\documentclass[intlimits,twoside,a4paper]{article}

\usepackage{amsmath,amssymb}
\usepackage{graphicx}

\usepackage[T2A]{fontenc}
\usepackage[cp1251]{inputenc}

\usepackage[eqsecnum]{cmpj2}

\issue{2015}{18}{4}{43005}
\doinumber{10.5488/CMP.18.43005}

\title[The self-consistent field model for Fermi systems]%
{The self-consistent field model for Fermi systems with account of three-body interactions %
}%
\author[Yu.M. Poluektov, A.A. Soroka, S.N. Shulga]
{Yu.M. Poluektov, A.A. Soroka, S.N. Shulga}%
\address{Akhiezer Institute for Theoretical Physics, %
National Science Center ``Kharkiv Institute of Physics and Technology'',
1 Akademichna St., 61108 Kharkiv, Ukraine} %

\date{Received August 3, 2015, in final form October 7, 2015}
\authorcopyright{Yu.M. Poluektov, A.A. Soroka, S.N. Shulga, 2015}

\begin{document}

\maketitle

\begin{abstract}
On the basis of a microscopic model of self-consistent field, the
thermodynamics of the many-particle Fermi system at finite
temperatures with account of three-body interactions is built and
the quasiparticle equations of motion are obtained. %
It is shown that the delta-like three-body interaction gives no
contribution into the self-consistent field, and the description of
three-body forces requires their nonlocality to be taken into account. %
The spatially uniform system is considered in detail, and on the
basis of the developed microscopic approach general formulas are
derived for the fermion's effective mass and the system's equation
of state with account of contribution from three-body forces. %
The effective mass and pressure are numerically calculated for the
potential of ``semi-transparent sphere'' type at zero temperature.
Expansions of the effective mass and pressure in powers of density
are obtained. It is shown that, with account of only pair forces,
the interaction of repulsive character reduces the quasiparticle
effective mass relative to the mass of a free particle, and the
attractive interaction raises the effective mass.
The question of thermodynamic stability of the Fermi system is
considered and the three-body repulsive interaction is shown to
extend the region of stability of the system with the interparticle
pair attraction. The quasiparticle energy spectrum is calculated
with account of three-body forces.

\pacs 05.30.Ch, 05.30.Fk, 05.70.-a
\end{abstract}

\section{Introduction}

The self-consistent field model is an effective approach for
describing systems of a large number of particles, even in the case
when the interaction between particles cannot be considered to be
weak and the density of a system to be low. This model is applicable
both for systems with a finite number of particles, such as atomic
nuclei \cite{Kirzhnits,Eisenberg,BBIG} or electronic shells of atoms
and molecules \cite{Hartree,Slater}, as well as for many-particle
systems, when describing them by methods of statistical physics.
The method of self-consistent field is especially useful in
studying spatially non-uniform systems and phase transitions
\cite{Braut,VLP}. The self-consistent field model serves as an
efficient main approximation in constructing a perturbation theory,
including such a theory for spatially non-uniform systems
\cite{Kirzhnits} and many-particle systems with broken symmetries
\cite{P1,P2,P3,P4}.

Systems of a large number of particles, obeying Fermi statistics,
within a phenomenological approach at large enough densities are
described in the language of the Fermi liquid theory
\cite{Landau,PN}, which originally was developed for normal limitless
systems. Afterwards, the Fermi liquid theory was generalized both
for systems of finite dimensions \cite{Migdal,Kolomiets} and for
many-particle Fermi systems with broken phase symmetry possessing
superfluid and superconducting properties \cite{AKPY}. The
Fermi-liquid approach is, in essence, a phenomenological variant of
the self-consistent field theory \cite{P5,P6}.

In contrast to the Fermi-liquid approach, in which the interaction
of quasiparticles is described phenomenologically by means of
interaction amplitudes, the account for the interaction between
real particles is laid in the basis of the developed microscopic approach. %
Usually the interaction between particles is described by means of
pair potentials, on the assumption that the presence of other
particles does not influence the interaction between the two
selected particles. Meanwhile, for particles possessing an internal
structure, the interaction between a pair of particles is changed due
to the presence of a third particle, that can be taken into account
by introducing potentials depending on the coordinates of three
particles. Such a representation follows from the consideration of
exchange and multipole interactions of more than two particles in
different orders of the perturbation theory \cite{AT,BS,SS}.
Essential is the fact that contribution from three-body forces not
only gives quantitative corrections to characteristics of a system
calculated with account of only pair interactions, but can be
necessary for qualitative understanding of some effects.

Accounting for three-body interactions is also important in the
theory of nuclear forces \cite{BBIG,Kolomiets,Bethe}, because they
model the dependence of the nucleon-nucleon interaction potential on
density. In particular, in the case of the interaction proposed by
Skyrme \cite{Skyrme,VB}, three-body interactions are described by a
simple delta-like potential. The role of three-body forces within
the framework of the quantum field approach is discussed in
reference~\cite{Hammer}.

The effects of three-body interactions should also manifest themselves in the interaction of structureless particles located in a
polarizable medium, for example electrons in a lattice.
However, this issue  is  completely unexplored so far.

In this paper, self-consistent field equations are obtained for
normal (non-superfluid) Fermi systems at finite temperatures with
account of three-body forces within the approach developed earlier
for Fermi systems with pair interactions \cite{P2,P3,P5,P6}. %
In general, the mean (or self-consistent) field approximation can be
formulated in different ways and, accordingly, different variants of
this approximation are possible. In most cases this approximation
is introduced at the level of equations of motion. %
In the statistical description of many-particle systems %
it is convenient to introduce the self-consistent field model, most
consistently and in most general form, on the level of Hamiltonian rather than on the level of equations
of motion. This permits in a natural
way not only to derive the self-consistent equations of motion, but
also to build the thermodynamics of a many-particle system already within
the scope of a particular model.

If the self-consistent field model is formulated in a way such that
all thermodynamic relations hold exactly in it, as it
takes place for the ideal quantum gases \cite{LL}, then the
structure of the model and all its parameters will be determined
uniquely. The method of constructing a model within the Hamiltonian
approach consists in splitting the exact initial Hamiltonian
into a sum of an approximating Hamiltonian and a residual
correlation Hamiltonian. The approximating (or self-consistent)
Hamiltonian is chosen as a quadratic operator form of most general
kind, that contains indeterminate fields at the initial stage. %
Such a Hamiltonian, describing a many-particle system in the
self-consistent field approximation, is reduced to a diagonal form
and the fields entering it are determined from the variational
principle. Within this description, the concept of quasiparticles
naturally arises in the microscopic approach. The effects associated
with the interaction of quasiparticles are accounted for by the
correlation Hamiltonian and can be calculated using the perturbation
theory. It is worth noting that this method is so general that it
can be also used for a consistent description of normal and
superfluid Bose systems \cite{P7,P8}, in particular phonons
\cite{P9}, as well as relativistic quantized fields with broken
symmetries \cite{P10,P11}.

From the derived self-consistent equations accounting for both pair
and three-body interactions, it follows that the delta-like
three-body interaction gives no contribution into the
self-consistent field. Therefore, a description of three-body
interactions requires accounting for their nonlocality. %
Thermodynamic properties of the spatially uniform
system are studied more in detail in the paper, and general formulae are derived
for the quasiparticle effective mass and the system's equation of
state at finite temperatures with account of both pair and
three-body interactions. It is shown that, with account of only pair
forces, the repulsive interaction reduces the quasiparticle
effective mass relative to the mass of a free particle, and the
attractive interaction raises it.

The effective mass of fermions and the pressure of system are
numerically calculated for the ``semi-transparent sphere'' potential
at zero temperature, and expansions of these quantities in powers of
density with account of three-body interactions are obtained. %
The influence of three-body forces on the stability of the
many-particle Fermi system is considered and it is shown that
accounting for three-body forces of repulsive character extends the
region of stability and can lead to stabilization of the system with
pair attraction between particles. The quasiparticle energy spectrum
is calculated with account of three-body forces.

\clearpage

\section{Hamiltonian of the Fermi system with account of three-body \\ interactions}

Potential energy of a system of $N$ particles possessing an internal
structure can be represented as a sum of pair, three-body, etc.
interactions
\begin{equation} \label{EQ01}
\begin{array}{l}
\displaystyle{%
  U({\bf r}_1,{\bf r}_2,\dots,{\bf r}_N)=\sum_{i<j}U({\bf r}_i,{\bf r}_j)+%
  \sum_{i<j<k}U({\bf r}_i,{\bf r}_j,{\bf r}_k)+\dots,
}%
\end{array}
\end{equation}
where $U({\bf r}_i,{\bf r}_j)=U({\bf r}_j,{\bf r}_i)$, $U({\bf
r}_i,{\bf r}_j,{\bf r}_k)$ is a symmetric function in all permutations of its
coordinates. In the second quantization representation, the
Hamiltonian of the many-particle system with account of pair and
tree-body interactions has the form
\begin{equation} \label{EQ02}
\begin{array}{l}
\displaystyle{%
  H=T+V_2+V_3\,.
}%
\end{array}
\end{equation}
Here,
\begin{equation} \label{EQ03}
\begin{array}{ll}
\displaystyle{%
  T=\int\!\! \rd q \rd q'\,\Psi^+(q)H_0(q,q')\Psi(q')
}%
\end{array}
\end{equation}
is the kinetic energy and the energy in external field $U_0(q)$, and
\begin{equation} \label{EQ04}
\begin{array}{c}
\displaystyle{%
  H_0(q,q')=-\frac{\hbar^2}{2m}\Delta\,\delta(q-q')+U_0(q)\,\delta(q-q').  %
}%
\end{array}
\end{equation}
The energies of pair $V_2$ and three-body $V_3$ interactions can be
written in the form
\begin{equation} \label{EQ05}
\begin{array}{ll}
\displaystyle{%
  V_2=\frac{1}{2!}\int\!\! \rd q \rd q'\,\Psi^+(q)\Psi^+(q')U(q,q')\Psi(q')\Psi(q), %
}
\end{array}
\end{equation}
\begin{equation} \label{EQ06}
\begin{array}{ll}
\displaystyle{%
  V_3=\frac{1}{3!}\int\!\! \rd q \rd q' \rd q''\,\Psi^+(q)\Psi^+(q')\Psi^+(q'')U(q,q',q'')\Psi(q'')\Psi(q')\Psi(q). %
}
\end{array}
\end{equation}
In the formulae (\ref{EQ03})--(\ref{EQ06}) and below, the symbol
$q\equiv ({\bf r},\sigma)$ designates the space coordinate ${\bf r}$
and the spin projection $\sigma$. We assume $s=1/2$. The field
operators obey the known anticommutation relations \cite{LL}. In what
follows, we consider the case when all the potentials depend
only on the space coordinates:
$U(q,q')=U({\bf r},{\bf r}')$,
$U(q,q',q'')=U({\bf r},{\bf r}',{\bf r}'')$. Besides that, we assume
that external potential does not depend on the spin projection, so
that $U_0(q)=U_0({\bf r})$. In order not to take into account the
condition of conservation of the total number of particles in all
computations, we assume that the considered system can exchange
particles with a thermostat. To account for it, a term with the
chemical potential $\mu$ is introduced into Hamiltonian, so that
$H_0(q,q')$ in (\ref{EQ03}) is replaced by one-particle Hamiltonian
\begin{equation} \label{EQ07}
\begin{array}{ll}
\displaystyle{%
   H(q,q')=H_0(q,q')-\mu\,\delta(q-q').
}%
\end{array}
\end{equation}
In what follows, when talking about Hamiltonian, we imply that it
includes the term with the chemical potential.

\section{The self-consistent field model with account of three-body \\
interactions} %

In order to proceed to the description of a many-particle Fermi system
within the self-consistent field model, let us represent the total
Hamiltonian (\ref{EQ02}) as a sum of two terms, i.e., the self-consistent
Hamiltonian $H_0$ and the correlation Hamiltonian $H_\textrm{C}$:
\begin{equation} \label{EQ08}
\begin{array}{l}
\displaystyle{%
  H=H_0+H_\textrm{C}. %
}
\end{array}
\end{equation}
The self-consistent Hamiltonian is defined by the relation
\begin{equation} \label{EQ09}
\begin{array}{ll}
\displaystyle{%
  H_0=\int\!\! \rd q \rd q'\,\Psi^+(q)\big[H(q,q')+W(q,q')\big]\Psi(q')+E_0,
} %
\end{array}
\end{equation}
which is quadratic in the field operators of creation and
annihilation. Equation (\ref{EQ09}) includes the self-consistent
potential $W(q,q')$, representing the mean field acting on a single
particle, as well as the non-operator term $E_0$, both still
indeterminate. Owing to hermiticity of the Hamiltonian, the next
property holds $W(q,q')=W^*(q',q)$. Note that taking account of the
non-operator term in (\ref{EQ09}) is essential for a consistent
description of the thermodynamics of a system within the
considered approach. The correlation Hamiltonian accounts for all
the effects, not accounted for in the self-consistent field model:
\begin{eqnarray} \label{EQ10}
H_\textrm{C}&=&\frac{1}{2!}\int\!\! \rd q \rd q' \Psi^+(q)\Psi^+(q')U({\bf r},{\bf r}')\Psi(q')\Psi(q) \nonumber\\
&& +\frac{1}{3!}\int\!\! \rd q \rd q' \rd q''\,\Psi^+(q)\Psi^+(q')\Psi^+(q'')U({\bf r},{\bf r}',{\bf r}'')\Psi(q'')\Psi(q')\Psi(q) \nonumber\\
&& -\int\!\! \rd q \rd q'\,\Psi^+(q)W(q,q')\Psi(q') - E_0.
\end{eqnarray}
Obviously, the total Hamiltonian $H$ has not changed in consequence
of the performed decomposition. Transition to the self-consistent
field model consists in describing the system using the approximate
quadratic Hamiltonian (\ref{EQ09}) instead of the exact Hamiltonian
(\ref{EQ08}). Entering (\ref{EQ09}), so
far indeterminate, quantities $W(q,q')$ and $E_0$ should be chosen
in an optimal manner, as it will be done below. The effects conditioned
by the correlation Hamiltonian can be accounted for by the
perturbation theory \cite{Kirzhnits,P2,P3,P4}. In this paper we
confine ourselves to consideration of only the main approximation.
Note also that in constructing the self-consistent field theory, we do not take into account the effects connected with
breaking the phase symmetry and those leading to the properties of
superfluidity and superconductivity~\cite{P2,P3}.

Since Hamiltonian (\ref{EQ09}) is quadratic in the field operators,
it can be represented in the form of Hamiltonian of an ideal gas of
quasiparticles. The field operators and the operators of creation
and annihilation of free particles $a_j^+$, $a_j$ are connected by the
relations
\begin{equation} \label{EQ11}
\begin{array}{ll}
\displaystyle{%
  \Psi(q)=\sum_j \phi_j^{(0)}\!(q)\,a_j, \qquad \Psi^+(q)=\sum_j
  \phi_j^{(0)*}\!(q)\,a_j^+,
} %
\end{array}
\end{equation}
where the functions $\phi_j^{(0)}\!(q)$ are solutions of the
Schr\"{o}dinger equation for free particles
\begin{equation} \label{EQ12}
\begin{array}{l}
\displaystyle{%
  \int\! \rd q' H_0(q,q')\phi_j^{(0)}\!(q')=\varepsilon_j^{(0)}\phi_j^{(0)}\!(q). %
}%
\end{array}
\end{equation}
Here, $j\equiv (\nu,\sigma)$, where $\nu$ is a full set of quantum
numbers describing the state of a particle except the spin
projection $\sigma$. To represent Hamiltonian (\ref{EQ09}) in the
form analogous to that for free particles, let us introduce the
quasiparticle operators $\gamma_j^+$, $\gamma_j$ connected with the
field operators (\ref{EQ11}) by the relations
\begin{equation} \label{EQ13}
\begin{array}{ll}
\displaystyle{%
  \Psi(q)=\sum_j \phi_j(q)\,\gamma_j, \quad \Psi^+(q)=\sum_j
  \phi_j^{*}(q)\,\gamma_j^+,
} %
\end{array}
\end{equation}
where the functions $\phi_j(q)$ are now solutions of the
self-consistent equation
\begin{equation} \label{EQ14}
\begin{array}{l}
\displaystyle{%
  \int\! \rd q' \big[ H(q,q')+W(q,q')\big]\phi_j(q')=\varepsilon_j \phi_j(q). %
}%
\end{array}
\end{equation}
Note that for $\phi_j(q)$ functions, the same as for
$\phi_j^{(0)}\!(q)$, the conditions of orthonormality and
completeness hold
\begin{equation} \label{EQ15}
\begin{array}{ll}
\displaystyle{%
  \int\! \rd q\, \phi_j^{*}(q)\phi_{j'}(q)=\delta_{jj'}, \qquad %
  \sum_j \phi_j^{*}(q)\phi_{j}(q')=\delta(q-q').
} %
\end{array}
\end{equation}
As a result, the self-consistent Hamiltonian (\ref{EQ09}) acquires
the form of the Hamiltonian of an ideal gas of quasiparticles
\begin{equation} \label{EQ16}
\begin{array}{l}
\displaystyle{%
  H_0= \sum_j \varepsilon_j\,\gamma_j^+\gamma_j + E_0, %
}
\end{array}
\end{equation}
where $\varepsilon_j$ means the quasiparticle energy
and $E_0$ means the energy of the background on which the
quasiparticles exist. In the self-consistent field model, the
reference of the energy cannot be arbitrary but, as will be shown
below, should be chosen in a way such that the thermodynamic
relations are satisfied. From (\ref{EQ11}) and (\ref{EQ13}) it
follows that the quasiparticle operators are explicitly expressed
through the operators of free particles
\begin{equation} \label{EQ17}
\begin{array}{ll}
\displaystyle{%
  \gamma_j=\sum_{j'}a_{j'}\int\!\rd q\phi_j^{*}(q)\phi_{j'}^{(0)}\!(q), \qquad %
  \gamma_j^+=\sum_{j'}a_{j'}^+\int\!\rd q\phi_j(q)\phi_{j'}^{(0)*}\!(q).
} %
\end{array}
\end{equation}
In a spatially uniform non-magnetic state, the operators of particles
and quasiparticles coincide, which corresponds to the known reasoning
in the Fermi liquid theory regarding the invariance of classification of
states during adiabatic ``switching-on'' of the interaction
\cite{Landau,PN}.

\section{Derivation of the self-consistent potential} %

Let us define the statistical operator
\begin{equation} \label{EQ18}
\begin{array}{ll}
\displaystyle{%
  \hat{\rho}_0=\exp\beta(\Omega-H_0),
} %
\end{array}
\end{equation}
where $\beta=1/T$ is the inverse temperature, the constant
$\Omega=-T\ln\!\big[\textrm{Sp}\,\re^{-\beta H_0}\big]$ is
determined from the normality condition $\textrm{Sp}\hat{\rho}_0=1$
and, as will be seen below, has the meaning of the thermodynamic
potential of the system in the self-consistent field model. The
average of an arbitrary operator $A$ in this approximation is
defined by the relation
\begin{equation} \label{EQ19}
\begin{array}{ll}
\displaystyle{%
  \langle A \rangle \equiv \textrm{Sp}(\hat{\rho}_0 A).
} %
\end{array}
\end{equation}
Since Hamiltonian (\ref{EQ16}) is quadratic, then for the averages
with the statistical operator (\ref{EQ18}) the Bloch-de
Dominicis (Wick) theorem \cite{BB} holds.

The ground state energy in the self-consistent field model is
determined from the requirement of equality of the averages
calculated with the statistical operator (\ref{EQ18}) for the exact
(\ref{EQ08}) and for the approximating (\ref{EQ09}) Hamiltonians:
\begin{equation} \label{EQ22}
\begin{array}{ll}
\displaystyle{%
  \langle H \rangle =  \langle H_0 \rangle.
} %
\end{array}
\end{equation}
Hence, we have
\begin{eqnarray} \label{EQ23}
E_0&=&\frac{1}{2!}\int\!\! \rd q \rd q'\,U({\bf r},{\bf r}') \big\langle\Psi^+\!(q)\Psi^+\!(q')\Psi(q')\Psi(q)\big\rangle  \nonumber\\
&&+\frac{1}{3!}\int\!\! \rd q \rd q' \rd q''\,U({\bf r},{\bf r}',{\bf r}'')\big\langle\Psi^+\!(q)\Psi^+\!(q')\Psi^+\!(q'')\Psi(q'')\Psi(q')\Psi(q)\big\rangle \nonumber\\
&& - \int\!\! \rd q \rd q'\, W(q,q')\big\langle\Psi^+\!(q)\Psi(q')\big\rangle.
\end{eqnarray}
Let us define the one-particle density matrix by the relation
\begin{equation} \label{EQ24}
\begin{array}{ll}
\displaystyle{%
  \rho(q,q')=\big\langle\Psi^+\!(q')\Psi(q)\big\rangle = \sum_i\phi_i(q)\phi_i^*\!(q')f_i. %
} %
\end{array}
\end{equation}
Here, the quasiparticle distribution function is defined by the expression %
$\displaystyle{\big\langle\gamma_i^+\gamma_j\big\rangle = f_i\,\delta_{ij}}$. %
Based on the form of Hamiltonian (\ref{EQ16}), a straightforward
calculation gives the Fermi-type distribution function
\begin{equation} \label{EQ25}
\begin{array}{ll}
\displaystyle{%
  f_i=f(\varepsilon_i)=\big[\!\exp(\beta\varepsilon_i)+1\big]^{-1}.  %
}
\end{array}
\end{equation}
Since the quasiparticle energy $\varepsilon_i$ is the functional of
$f_i$, the formula (\ref{EQ25}) represents a complicated nonlinear
equation for the distribution function, being similar to that which
takes place in the Landau phenomenological theory of a Fermi liquid \cite{Landau}. %
The energy $E_0$ expressed through $\rho(q,q')$ has the form:
\begin{eqnarray} \label{EQ26}
E_0&=&\frac{1}{2!}\int\!\! \rd q \rd q'\,U({\bf r},{\bf r}') \big[\rho(q,q)\rho(q',q')-\rho(q',q)\rho(q,q')\big]  \nonumber\\
&&+\frac{1}{3!}\int\!\! \rd q \rd q' \rd q''\,U({\bf r},{\bf r}',{\bf r}'')
   \Big[2\rho(q,q')\rho(q',q'')\rho(q'',q) \nonumber\\
&&-\,3\rho(q,q)\rho(q',q'')\rho(q'',q')+\rho(q,q)\rho(q',q')\rho(q'',q'')\Big]\nonumber\\
&&-\int\!\! \rd q \rd q'\, W(q,q')\rho(q',q).
\end{eqnarray}

The variation of the thermodynamic potential %
$\Omega=-T\ln\!\big[\!\sum_n\!\big\langle n\big|\re^{-\beta H_0}\big|n\big\rangle \big]$ %
is equal to the averaged variation of Hamiltonian (\ref{EQ09}): %
\begin{equation} \label{EQ27}
\begin{array}{ll}
\displaystyle{%
  \delta\Omega=\frac{\sum_n\!\big\langle n\big|\re^{-\beta H_0}\delta H_0\big|n\big\rangle}
               {\sum_n\!\big\langle n\big|\re^{-\beta H_0}\big|n\big\rangle} =  
  \big\langle\delta H_0\big\rangle.
}
\end{array}
\end{equation}
From the requirement that the variation of the thermodynamic
potential with respect to the density matrix
(\ref{EQ24}) vanishes $\delta\Omega\big/\delta\rho(q,q')=0,$ %
we obtain the expression for the self-consistent potential which
consists of the contributions from pair and three-body interactions
$W(q,q')=W^{(2)}\!(q,q')+W^{(3)}\!(q,q')$, where
\begin{equation} \label{EQ28}
  W^{(2)}\!(q,q')=-U({\bf r},{\bf r}')\rho(q,q')+\delta(q-q')\int\!\! \rd q''U({\bf r},{\bf r}'')\rho(q'',q''), %
\end{equation}
\begin{eqnarray} \label{EQ29}
W^{(3)}\!(q,q')\!\!\!&=&\!\!\!\int\!\! \rd q''\,U({\bf r},{\bf r}',{\bf r}'')\big[\rho(q,q'')\rho(q'',q') - \rho(q,q')\rho(q'',q'')\big] \nonumber \\
&&\!\!\! + \frac{1}{2}\delta(q-q')\int\!\! \rd q''\rd q'''\,U({\bf r},{\bf r}'',{\bf r}''')\big[\rho(q'',q'')\rho(q''',q''') - \rho(q'',q''')\rho(q''',q'')\big]. \qquad %
\end{eqnarray}
The self-consistent field can be also derived by an equivalent
method from the requirement that the variation of the functional %
$I\equiv\big\langle H-H_0\big\rangle$ with respect to the density
matrix vanishes, where $W(q,q')$ and the parameter $E_0$ are not
varied \cite{P2,P3}. Such variation rule is conditioned by the fact
that, as may be checked, the variations of the functional $\delta I$
with respect to $\delta W(q,q')$ and $\delta E_0$ are mutually
compensated. Equation (\ref{EQ14}), together with the derived
potentials (\ref{EQ28}), (\ref{EQ29}), enables us to obtain the
quasiparticle wave functions $\phi_i(q)$ and the quasiparticle
energies $\varepsilon_i$:
\begin{equation} \label{EQ30}
  \varepsilon_i= -\frac{\hbar^2}{2m}\int\!\! \rd q\,\phi_i^*\!(q)\Delta\phi_i(q) + %
  \int\!\! \rd q\, U_0({\bf r})\big|\phi_i(q)\big|^2 + \int\!\! \rd q \rd q'\,W(q,q')\phi_i^*\!(q)\phi_i(q') - \mu.
\end{equation}
Substitution of the potentials (\ref{EQ28}), (\ref{EQ29}) into
equation (\ref{EQ14}) leads to the integro-differential equation
\begin{eqnarray} \label{EQ31}
&&-\frac{\hbar^2}{2m}\Delta\phi_i(q)+\big[U_0({\bf r})-\mu-\varepsilon_i\big]\phi_i(q) \nonumber\\%
&& + \left\{ \!\int\!\! \rd q'\,U({\bf r},{\bf r}')\rho(q',q')
+ \frac{1}{2}\int\!\! \rd q' \rd q''\,U({\bf r},{\bf r}',{\bf r}'')\big[\rho(q',q')\rho(q'',q'')-\rho(q',q'')\rho(q'',q')\big]\right\}\phi_i(q)  \nonumber\\
&& + \int\!\! \rd q'\rd q''\,U({\bf r},{\bf r}',{\bf r}'')\big[\rho(q,q'')\rho(q'',q')-\rho(q,q')\rho(q'',q'')\big]\phi_i(q') \nonumber\\
&& - \int\!\! \rd q' \,U({\bf r},{\bf r}')\rho(q,q')\phi_i(q') = 0.
\end{eqnarray}
The chemical potential $\mu$ is associated with the average number
of particles $N$ by the relation %
\begin{equation} \label{EQ32}
\begin{array}{ll}
\displaystyle{%
  N= \int\!\! \rd{\bf r}\,n({\bf r}), \qquad  n({\bf r})=\sum_\sigma\rho(q,q). %
}
\end{array}
\end{equation}

In many cases, finding the equilibrium characteristics of the
researched system does not require the calculation of the quasiparticle
wave functions, but it is sufficient to know the one-particle
density matrix. From equations (\ref{EQ24}) and (\ref{EQ31}), the
equation for the one-particle density matrix follows
\begin{eqnarray} \label{EQ33}
&& \hspace{-1.3cm}  \frac{\hbar^2}{2m} \big[\Delta\rho(q,q')-\Delta'\rho(q,q')\big]-\!\big[U_0({\bf r})-U_0({\bf r}')\big]\rho(q,q') \nonumber \\
&& \hspace{-1.3cm}  +\int\!\! \rd q''\big[ U({\bf r},{\bf r}'')-U({\bf r}',{\bf r}'')\big]\!
  \big[\rho(q,q'')\rho(q'',q')-\rho(q'',q'')\rho(q,q')\big]  \nonumber \\
&& \hspace{-1.3cm}  + \int\!\! \rd q''\rd q'''\big[U({\bf r},{\bf r}'',{\bf r}''')-U({\bf r}',{\bf r}'',{\bf r}''')\big]\!  \big[\rho(q,q''')\rho(q''',q'')-\rho(q,q'')\rho(q''',q''')\big]\rho(q'',q') \nonumber \\
&& \hspace{-1.3cm}  +\frac{1}{2}\,\rho(q,q')\int\!\! \rd q''\rd q'''\big[U({\bf r},{\bf r}'',{\bf r}''')-U({\bf r}',{\bf r}'',{\bf r}''')\big]\! \big[\rho(q'',q'')\rho(q''',q''')-\rho(q'',q''')\rho(q''',q'')\big]  = 0.
\end{eqnarray}

In the absence of magnetic effects %
$\rho(q,q')=\rho({\bf r},{\bf r}')\delta_{\sigma\sigma'}$, %
and the self-consistent potential is diagonal in spin indices as well %
$W(q,q')=[W^{(2)}\!({\bf r},{\bf r}')+W^{(3)}\!({\bf r},{\bf r}')]\delta_{\sigma\sigma'}$, %
where
\begin{eqnarray} \label{EQ34}
  W^{(2)}\!({\bf r},{\bf r}')&=&-U({\bf r},{\bf r}')\rho({\bf r},{\bf r}')+
  2\,\delta({\bf r}-{\bf r}')\int\!\! \rd{\bf r}''U({\bf r},{\bf r}'')\rho({\bf r}'',{\bf r}''), \\
%
\label{EQ35}
W^{(3)}\!({\bf r},{\bf r}')&=&\int\!\! \rd{\bf r}''\,U({\bf r},{\bf r}',{\bf r}'') \rho({\bf r},{\bf r}'')\rho({\bf r}'',{\bf r}')
- 2\,\rho({\bf r},{\bf r}')\int\!\! \rd{\bf r}''\,U({\bf r},{\bf r}',{\bf r}'')\rho({\bf r}'',{\bf r}'') \nonumber \\
&&  + \delta({\bf r}-{\bf r}')\int\!\! \rd{\bf r}''\rd{\bf r}'''\,U({\bf r},{\bf r}'',{\bf r}''')  \big[2\rho({\bf r}'',{\bf r}'')\rho({\bf r}''',{\bf r}''') - \rho({\bf r}'',{\bf r}''')\rho({\bf r}''',{\bf r}'')\big]. %
\end{eqnarray}
In what follows we consider the system without account of
magnetic effects.

\section{Thermodynamic relations}

Let us formulate the self-consistent field model at finite
temperatures in a way such that all the thermodynamic relations
would hold already in this approximation. This requirement leads to
a unique formulation of the model. Entering the definition of
the statistical operator quantity $\Omega$ has the meaning of the
grand thermodynamic potential in the self-consistent field model.
The entropy is defined through the statistical operator (\ref{EQ18})
by the known expression
\begin{equation} \label{EQ36}
\begin{array}{ll}
\displaystyle{%
  S = -\textrm{Sp}\big(\hat{\rho}_0\ln\hat{\rho}_0 \big).
} %
\end{array}
\end{equation}
With the help of (\ref{EQ36}) it is easy to verify that the usual
thermodynamic definition of the grand thermodynamic potential holds
$\Omega=E-TS-\mu N$, where $E$ is the total energy of the system. %
This potential is a function of the temperature $T$ and of the chemical
potential $\mu$, as well as a functional of the one-particle density
matrix $\rho(q,q')=\rho(q,q';T,\mu)$, which also depends on these
quantities. However, by virtue of the fact that the
self-consistent potential was derived from the condition %
$\delta\Omega\big/\delta\rho(q,q')=0$, when finding the derivatives
of $\Omega$ with respect to temperature and chemical potential, one
should account for only the explicit dependence of the
thermodynamic potential on these quantities. As a consequence, %
at a fixed volume of the system, the usual thermodynamic relation
proves to be fulfilled in the self-consistent field model:
\begin{equation} \label{EQ37}
  \rd\Omega = -S\rd T -N\rd\mu.
\end{equation}
Calculation of the entropy and the number of particles either by
means of the averaging with the density matrix (\ref{EQ18}) or with the
help of the thermodynamic relations
$S=-(\partial\Omega/\partial T)_\mu$, $N=-(\partial\Omega/\partial \mu)_T$ %
gives the same result. With the correct choice of the energy $E_0$, %
any kind of inconsistency in statistical description of systems
within the self-consistent field model, mentioned in
\cite{Kirzhnits}, does not appear.

The energy (\ref{EQ26}), with account of the obtained
self-consistent potential (\ref{EQ28}), (\ref{EQ29}), acquires the form %
\begin{eqnarray} \label{EQ38}
E_0\!\!\!&=\!\!\!&\frac{1}{2}\int\!\! \rd q \rd q'\,U({\bf r},{\bf r}') \big[\rho(q,q')\rho(q',q)-\rho(q,q)\rho(q',q')\big] \nonumber \\
&&  +\!\int\!\! \rd q \rd q'\!\rd q'' \!U({\bf r},{\bf r}',{\bf r}'')
   \!\!\bigg[\!\rho(q,q)\rho(q',q'')\rho(q'',q') \!-\! \frac{1}{3}\rho(q,q)\rho(q',q')\rho(q'',q'')\!\nonumber \\
&&   -\!\frac{2}{3}\rho(q,q')\rho(q',q'')\rho(q'',q)\!\bigg],
\end{eqnarray}
and the thermodynamic potential is determined by the formula
\begin{equation} \label{EQ39}
\begin{array}{ll}
\displaystyle{%
  \Omega = E_0 - T\sum_i\ln\!\left( 1+ \re^{-\beta\varepsilon_i}\right).
}
\end{array}
\end{equation}
The total energy can be calculated either by averaging the
Hamiltonian operator or by the use of the thermodynamic potential
(\ref{EQ39})
\begin{equation} \label{EQ40}
\begin{array}{ll}
\displaystyle{%
  E=\Omega - \mu\frac{\partial\Omega}{\partial\mu}-T\frac{\partial\Omega}{\partial T} %
}
\end{array}
\end{equation}
and has the form
\begin{eqnarray} \label{EQ41}
E\!\!\!&=\!\!\!&\sum_i\varepsilon_i f_i + \mu N +\frac{1}{2}\int\!\! \rd q \rd q'\,U({\bf r},{\bf r}') \big[\rho(q,q')\rho(q',q)-\rho(q,q)\rho(q',q')\big] \nonumber\\
&&  +\!\int\!\! \rd q \rd q'\!\rd q'' \!U({\bf r},{\bf r}',{\bf r}'')
\!\!\bigg[\!\rho(q,q)\rho(q',q'')\rho(q'',q') \!-\! \frac{1}{3}\rho(q,q)\rho(q',q')\rho(q'',q'')\nonumber\\
&& -\!\frac{2}{3}\rho(q,q')\rho(q',q'')\rho(q'',q)\!\bigg].
\end{eqnarray}

The entropy in this model is expressed through the distribution
function  formally in the same way as in the model of an ideal gas
\begin{equation} \label{EQ42}
\begin{array}{ll}
\displaystyle{%
 S = -\sum_i\!\big[f_i\ln f_i +(1-f_i)\ln(1-f_i)\big],
}
\end{array}
\end{equation}
but, as remarked previously, the distribution function itself is
derived from the complicated nonlinear equation (\ref{EQ25}) and
includes the effects conditioned by both pair and three-body
interactions.

The total number of particles (\ref{EQ32}) is expressed through the
distribution function by the formula
\begin{equation} \label{EQ43}
\begin{array}{ll}
\displaystyle{%
 N = \sum_i f_i\,.
}
\end{array}
\end{equation}
Since, owing to interaction, the energy of a single particle
(\ref{EQ30}) differs from the energy of a free particle and includes
collective effects, then in this case a particle should be treated
as a quasiparticle and the function $f_i$ as the distribution
function of quasiparticles. According to (\ref{EQ32}), (\ref{EQ43}),
the number of initial free particles and the number of
quasiparticles coincide, as it takes place in the Landau theory of a
normal Fermi liquid \cite{Landau}. In the presence of pair
correlations, leading to the property of superfluidity, the number
of quasiparticles is less than the number of initial free particles
\cite{P2,P3}, because some fraction of particles participates in the
formation of the pair condensate.

Within the framework of the microscopic model of self-consistent
field, formulae (\ref{EQ38})--(\ref{EQ43}) describe in general form the thermodynamics of the many-particle Fermi system at
finite temperatures with account of pair and three-body interactions. %

\section{Interaction potentials in the self-consistent field model}

In calculations within the self-consistent field approximation,
instead of realistic potentials by which individual particles
interact in the vacuum and that usually model a strong repulsion at
short distances, the effective potentials with a set of adjustable
parameters are used \cite{Skyrme,VB}. A delta-like potential is often
chosen as the simplest model potential
of the interparticle interaction. Such a choice of the pair
interaction potential %
$U({\bf r}',{\bf r}'')=t_2\delta({\bf r}'-{\bf r}'')$ %
results in the following form of the self-consistent field %
\begin{equation} \label{EQ44}
\begin{array}{ll}
\displaystyle{%
  W^{(2)}\!({\bf r},{\bf r}')= \frac{t_2}{2}\,\delta({\bf r}-{\bf r}')\,n({\bf r}), %
}%
\end{array}
\end{equation}
where $n({\bf r})=2\rho({\bf r},{\bf r})$ is the particle number
density. In this case, the contribution into the self-consistent
field (\ref{EQ34}) of the first (exchange) term proves to be two
times less by absolute value than the contribution of the second
term, describing the direct interaction, and has an opposite sign.

As can be immediately verified, when choosing three-body forces in
the delta-like form
\begin{equation} \label{EQ45}
\begin{array}{ll}
\displaystyle{%
  U({\bf r}',{\bf r}'',{\bf r}''')= t_3\,\delta({\bf r}'-{\bf r}'')\delta({\bf r}'-{\bf r}'''), %
}%
\end{array}
\end{equation}
which is used, for example, in \cite{Skyrme}, the self-consistent potential %
$W^{(3)}\!({\bf r},{\bf r}')$ (\ref{EQ35}) vanishes. %
Therefore, in order to obtain a nonvanishing contribution of
three-body forces into the self-consistent field, we have to take
into account their nonlocality, that is their finite radius of action. %
Physically this property is conditioned by the Fermi statistics, not
allowing for the presence of more than two particles with opposite spins
in the same space point. In the case of an arbitrary three-body
potential of the interparticle interaction, its contribution into
the self-consistent potential (\ref{EQ35}) can be represented in the
from
\begin{equation} \label{EQ46}
\begin{array}{ll}
\displaystyle{%
  W^{(3)}\!({\bf r},{\bf r}')=
  a_1\!({\bf r},{\bf r}')-2\rho({\bf r},{\bf r}') a_2\!({\bf r},{\bf r}') + %
  \delta({\bf r}-{\bf r}')\big[2b_1({\bf r})-b_2({\bf r})\big],
}%
\end{array}
\end{equation}
where
\begin{align} \label{EQ47}
& a_1\!({\bf r},{\bf r}')\equiv \int\!\! \rd{\bf r}''\,U({\bf r},{\bf r}',{\bf r}'')\rho({\bf r},{\bf r}'')\rho({\bf r}'',{\bf r}'),&
& a_2\!({\bf r},{\bf r}')\equiv \int\!\! \rd{\bf r}''\,U({\bf r},{\bf r}',{\bf r}'')\rho({\bf r}'',{\bf r}''), \nonumber\\
& b_1\!({\bf r})\equiv \int\!\! \rd{\bf r}'\rd{\bf r}''\,U({\bf r},{\bf r}',{\bf r}'')\rho({\bf r}',{\bf r}')\rho({\bf r}'',{\bf r}''), &
& b_2\!({\bf r})\equiv \int\!\! \rd{\bf r}'\rd{\bf r}''\,U({\bf r},{\bf r}',{\bf r}'')\rho({\bf r}',{\bf r}'')\rho({\bf r}'',{\bf r}').
\end{align}

Let us discuss the issue of choosing the potentials of
interparticle interactions, which are suitable for the
self-consistent field model. In a spatially uniform state, the
interaction potentials should satisfy two conditions. They must not
change under the replacement %
${\bf r} \rightarrow {\bf r}+{\bf a}$ (${\bf a}$ is an arbitrary vector) %
of all the vectors they depend on, and must be symmetric relative to any
permutation of their arguments. The pair potential satisfying these
conditions has the form
\begin{equation} \label{EQ48}
\begin{array}{ll}
\displaystyle{%
  U({\bf r},{\bf r}')= \frac{1}{2}\big[U_2({\bf r}-{\bf r}')+U_2({\bf r}'-{\bf r})\big], %
}%
\end{array}
\end{equation}
where $U_2({\bf r})$ is the function of a vector argument.
The three-body potential, depending on the pairs of vector differences
and satisfying the similar conditions, is as follows:
\begin{eqnarray} \label{EQ49}
U({\bf r},{\bf r}',{\bf r}'')&=& \frac{1}{6}\big[
U_3({\bf r}-{\bf r}',{\bf r}-{\bf r}'')+U_3({\bf r}-{\bf r}'',{\bf r}-{\bf r}')+U_3({\bf r}'-{\bf r},{\bf r}'-{\bf r}'')\nonumber \\
&& +\,U_3({\bf r}'-{\bf r}'',{\bf r}'-{\bf r})+U_3({\bf r}''-{\bf r},{\bf r}''-{\bf r}')+U_3({\bf r}''-{\bf r}',{\bf r}''-{\bf r}) \big].
\end{eqnarray}
Here, $U_3({\bf r},{\bf r}')$ is a function of two vector arguments. %
If there is a dependence only on the distances between particles,
and the function in (\ref{EQ49}) is symmetric %
$U_3({\bf r},{\bf r}')=U_3({\bf r}',{\bf r})$, then %
$U({\bf r},{\bf r}')=U_2(|{\bf r}-{\bf r}'|)$ and
\begin{equation} \label{EQ50}
\begin{array}{ll}
\displaystyle{%
  U({\bf r},{\bf r}',{\bf r}'')= \frac{1}{3}\big[
  U_3(|{\bf r}-{\bf r}'|,|{\bf r}-{\bf r}''|)+U_3(|{\bf r}'-{\bf r}|,|{\bf r}'-{\bf r}'')|+U_3(|{\bf r}''-{\bf r}|,|{\bf r}''-{\bf r}'|) %
\big]. %
}%
\end{array}
\end{equation}
In particular, the three-body potential can be chosen in the form
proposed in reference~\cite{SS}:
\begin{equation} \label{EQ51}
\begin{array}{ll}
\displaystyle{%
  U(|{\bf r}-{\bf r}'|,|{\bf r}-{\bf r}''|)=u_0\,\exp\!\left(-\frac{\,|{\bf r}-{\bf r}'|+|{\bf r}-{\bf r}''|}{r_0}\right). %
}%
\end{array}
\end{equation}
There is also a possibility of choosing the potential in the form of the Gauss function: %
\begin{equation} \label{EQ52}
\begin{array}{ll}
\displaystyle{%
  U(|{\bf r}-{\bf r}'|,|{\bf r}-{\bf r}''|)=\frac{u_0}{\pi^{3/2}r_0^3}\,\exp\!\left[-\frac{\,({\bf r}-{\bf r}')^2+({\bf r}-{\bf r}'')^2}{r_0^2}\right]. %
}%
\end{array}
\end{equation}
Such a choice is characteristic because in the limit %
$r_0\rightarrow 0$, the potential (\ref{EQ52}) turns into the
potential of zero radius (\ref{EQ45}), and its contribution into the
self-consistent field vanishes. In principle, the model three-body
potential can be chosen to depend on the three distances between
three particles
\begin{equation} \label{EQ53}
\begin{array}{ll}
\displaystyle{%
  U({\bf r},{\bf r}',{\bf r}'')= U_3(|{\bf r}-{\bf r}'|,|{\bf r}-{\bf r}''|,|{\bf r}'-{\bf r}''|). %
}%
\end{array}
\end{equation}
Note that the derivation from first principles of the potential of
interaction of three atoms as structureless entities presents a
complex quantum mechanical problem \cite{Sarry}.

The presently developed programs use different types of the model
interatomic three-body potentials. These potentials are often of
a complex form and in most cases include the angular dependence
\cite{Gale}, essential in the formation of the crystalline phase. In
our paper, which can be used, for example, for the description of
the system of $^3\rm{He}$ atoms in the gaseous or liquid phases, we
restrict ourselves to the use of potentials of the simplest form,
not depending on angles.

The potentials which rapidly tend to infinity at small distances are often used for describing the interaction between particles. An
example of such a potential is given by the Lennard-Jones potential %
\begin{equation} \label{EQ54}
\begin{array}{ll}
\displaystyle{%
  U_\textrm{LJ}(r)= 4\varepsilon\!\left[\left(\frac{\sigma}{r}\right)^{\!12}-\left(\frac{\sigma}{r}\right)^{\!6}\right], %
}%
\end{array}
\end{equation}
containing two parameters: the distance $\sigma$ and the energy
$\varepsilon$. The use of the potentials with the hard core leads to
considerable difficulties, especially in the quantum mechanical
description \cite{Thouless}. Such potentials do not have the Fourier
image, and the self-consistent field
becomes infinity when using them. Sometimes, this fact is used as justification of
non-applicability of the self-consistent field model in this or
that case, for example for describing the liquid \cite{Bazarov}. %
Thus, the issue of the choice of the interparticle interaction
potentials is of principal significance for problems of statistical
physics. The use of the potential which rapidly tends to infinity at
small distances means that an atom or another compound particle
retains its identity at arbitrary high pressures. Meanwhile, it is
clear that the critical pressure should exist at which the atoms approach
each other so closely that they should be ``crushed'' and loose
their identity. Therefore, the requirement of absolute
impermeability of particles at arbitrary high pressures is
excessively rigorous and unphysical and, in our opinion, it is more
reasonable to use the potentials which tend to a finite value at
small distances. An example of such a potential is given by the known
Morse potential
$U_\textrm{M}(r)=\varepsilon\big\{\!\exp[-2(r-r_0)/a]-2\exp[-(r-r_0)/a]\big\}$, %
where $\varepsilon$ is the energy parameter and $r_0,a$ are specific
distances. Notice that, from the quantum mechanical point of view,
the use of bounded potentials means the possibility for particles to
tunnel through each other with some probability. %
The problem of calculating the integral $\int\!U(r)\rd{\bf r}$
diverging for the potentials of type (\ref{EQ54}) is encountered,
for example, when deriving the known Gross-Pitaevskii equation,
which is presently  widely used in describing atomic Bose-Einstein
condensates \cite{PS}. Here, the diverging integral is replaced by a
finite value coinciding with the scattering length under the
fulfilment of the Born approximation, which in essence means the use
of the potential not diverging rapidly to infinity at small distances. %
It should also be noted that quantum chemical calculations lead
to the potentials having a finite value of energy at zero \cite{AS,ATB}. %

The model of ``semi-transparent sphere'' potential is
quite often used as the simplest form of the pair potential where the noted
problems with divergences are absent:
\begin{equation} \label{EQ55}
 U_2(r)=\left\{
               \begin{array}{ll}
                 U_{2\rm{m}}, & r<r_2, \\
                 \,\,0, & r>r_2.
               \end{array}
               \right.
\end{equation}
This potential is determined by two parameters, one of which
$U_{2\rm{m}}$ has the energy dimension and defines the interaction
strength, and the second one $r_2$ of the length dimension defines
the radius of interaction. Such an observable quantity as the
scattering length in the Born approximation (which, as should be
noted, is far from being always valid for realistic potentials) can be
expressed through the parameters of the
potential (\ref{EQ55}): %
$a_0=(m/4\pi\hbar^2)\int U_2(r)\rd{\bf r} = mU_{2\rm{m}}r_{2}^3\big/3\hbar^2$. %
A similar model can be also used in the case of three-body
forces. For the potential (\ref{EQ50}) depending on pairs of
distances between particles, we have
\begin{equation} \label{EQ56}
 U_3(r,r')=\left\{
               \begin{array}{ll}
                 U_{3\rm{m}},& r<r_3,\,r'<r_3, \\
                 \,\,0,   & \textrm{else}.
               \end{array} \right.
\end{equation}
Here, there are also two parameters: the strength $ U_{3\rm{m}}$ and
the radius $r_3$. For this choice, the total potential %
$U({\bf r},{\bf r}',{\bf r}'')= U_{3\rm{m}}$, if the distances between
each pair of three particles are less than $r_3$. However, if two
distances between particles are less than $r_3$, and the third one is
greater than $r_3$ and less than $2r_3$, then
$U({\bf r},{\bf r}',{\bf r}'')= U_{3\rm{m}}/3$. %
In other cases, the potential vanishes.

For the  potential (\ref{EQ53}), depending on three distances
between particles, the model of ``semi-trans\-parent sphere''
potential is defined by the formula
\begin{equation} \label{EQ57}
 U_3(r,r',r'')=\left\{
               \begin{array}{ll}
                  U_{3\rm{m}}, & r<r_3, \ r'<r_3, \ r''<r_3, \\
                 \,\,0,   & \textrm{else}.
               \end{array} \right.
\end{equation}
In this case, the potential is nonzero only under the condition that
the distances between each pair of three particles are less than $r_3$. %
Let us consider  more in detail the spatially uniform system of Fermi
particles in the absence of an external field and with account of
three-body forces.

\section{The spatially uniform system}

In the spatially uniform state, the one-particle density matrix is a
function of absolute value of the coordinate difference %
$\rho({\bf r},{\bf r}')=\rho(|{\bf r}-{\bf r}')|$, and the particle
number density $n=2\rho(0)$ is constant. For the potential
(\ref{EQ50}) depending on pairs of distances between particles, the
coefficients (\ref{EQ47}) which determine the contribution of
three-body forces into the self-consistent potential acquire the form %
\begin{eqnarray} \label{EQ58}
a_1\!({\bf r}-{\bf r}')&\!\!\!=\!\!\!&\frac{2}{3} \int\!\! \rd{\bf r}''\,U_3(|{\bf r}-{\bf r}'|,r'')\rho(r'')\rho(|{\bf r}-{\bf r}'+{\bf r}''|) + \frac{1}{3} \int\!\! \rd{\bf r}''\,U_3(|{\bf r}-{\bf r}'+{\bf r}''|,r'')\rho(r'')\rho(|{\bf r}-{\bf r}'+{\bf r}''|), \nonumber \\
a_2\!({\bf r}-{\bf r}')&\!\!\!=\!\!\!&\frac{2\rho(0)}{3} \int\!\! \rd{\bf r}''\,U_3(|{\bf r}-{\bf r}'|,r'') + \frac{\rho(0)}{3} \int\!\! \rd{\bf r}''\,U_3(|{\bf r}-{\bf r}'+{\bf r}''|,r''), \nonumber \\
b_1&\!\!\!=\!\!\!&\rho^2(0)\int\!\! \rd{\bf r}\rd{\bf r}'\,U_3(r,r'), \nonumber \\
b_2&\!\!\!=\!\!\!&\frac{1}{3}\int\!\! \rd{\bf r}\rd{\bf r}'\,U_3(|{\bf r}-{\bf r}'|,r')\rho^2(r) +
      \frac{2}{3}\int\!\! \rd{\bf r}\rd{\bf r}'\,U_3(r,r')\rho^2(r').
\end{eqnarray}
Expansions of the three-body potential and the one-particle density
matrix in Legendre polynomials can be used:
\begin{eqnarray} \label{EQ59}
U_3(|{\bf r}-{\bf r}'|,r'')&=&\sum_{l=0}^\infty U_{3l}(r,r';r'')P_l(\cos\theta),
\nonumber \\
\rho(|{\bf r}-{\bf r}'|)&=&\sum_{l=0}^\infty \rho_l(r,r')P_l(\cos\theta),
\end{eqnarray}
where
\begin{eqnarray} \label{EQ60}
U_{3l}(r,r';r'')&\!\!\!=\!\!\!&\frac{2l+1}{2}\int_{-1}^{1}U_3\!\left(\!\sqrt{r^2+r'^2-2rr'x},r''\right)
\!P_l(x)\,\rd x, \nonumber \\
\rho_l(r,r')&\!\!\!=\!\!\!&\frac{2l+1}{2}\int_{-1}^{1}\rho\!
\left(\!\sqrt{r^2+r'^2-2rr'x}\,\right)\!P_l(x)\,\rd x
=\frac{2l+1}{2\pi^2}\int_0^{\!\infty}\!f(\varepsilon_k)j_l(kr)j_l(kr')k^2\rd k,
\end{eqnarray}
$j_l(x)$ is the spherical Bessel function. With these expansions
taken into account we have:
\begin{eqnarray} \label{EQ61}
a_1\!(r)&=&\frac{8\pi}{3} \int_0^{\!\infty}\!\! U_3(r,r')\rho(r')\rho_0(r,r')r'^2\rd r' +
\frac{4\pi}{3} \sum_{l=0}^\infty \frac{1}{2l+1}\int_0^{\!\infty}\!\! U_{3l}(r,r';r')\rho(r')\rho_l(r,r')r'^2\rd r', \nonumber \\
a_2\!(r)&=&\frac{8\pi\rho(0)}{3} \int_0^{\!\infty}\!\! U_3(r,r')r'^2\rd r' +
\frac{4\pi\rho(0)}{3} \int_0^{\!\infty}\!\! U_{30}(r,r';r')r'^2\rd r', \nonumber \\
b_1&=&16\pi^2\rho^2(0) \int_0^{\!\infty}\!\! r^2\rd r \int_0^{\!\infty}\!\! U_3(r,r')r'^2\rd r', \nonumber \\
b_2&=&\frac{32\pi^2}{3} \int_0^{\!\infty}\!\! \rho^2(r)r^2\rd r \int_0^{\!\infty}\!\! U_3(r,r')r'^2\rd r' + \frac{16\pi^2}{3} \int_0^{\!\infty}\!\! \rho^2(r)r^2\rd r \int_0^{\!\infty}\!\! U_{30}(r,r';r')r'^2\rd r'.
\end{eqnarray}

In the spatially uniform case, the state of a single particle can be  
characterized by its momentum, and equation (\ref{EQ14}) admits
solutions in the form of plane waves:
\begin{equation} \label{EQ62}
\begin{array}{l}
\displaystyle{%
  \phi_j(q)=\frac{\delta_{\sigma\sigma'}}{\sqrt{V}}\re^{\ri{\bf k}{\bf r}}, %
}%
\end{array}
\end{equation}
where $q\equiv ({\bf r},\sigma)$, $j\equiv ({\bf k},\sigma')$. %
As was noted, in the absence of magnetic effects %
$\rho(q,q')=\delta_{\sigma\sigma'}\rho({\bf r}-{\bf r}')$, %
$W(q,q')=\delta_{\sigma\sigma'}W({\bf r}-{\bf r}')$. %
Equation (\ref{EQ30}) gives the quasiparticle energy not depending
on the spin projection: %
\begin{equation} \label{EQ63}
\begin{array}{l}
\displaystyle{%
  \varepsilon_{{\bf k}}=\frac{\hbar^2k^2}{2m} - \mu + W_{{\bf k}}, %
}%
\end{array}
\end{equation}
where $W_{{\bf k}}=\int\! \rd{\bf r}W({\bf r})\re^{\ri{\bf k}{\bf r}}$. %
The self-consistent potential can be represented as a sum of
direct and exchange terms
\begin{equation} \label{EQ64}
\begin{array}{l}
\displaystyle{%
  W({\bf r})=W_0\delta({\bf r})+W_\textrm{C}({\bf r}), %
}%
\end{array}
\end{equation}
and, if the interaction depends only on a distance between
particles, then also $W_\textrm{C}({\bf r})=W_\textrm{C}(r)$. In this case,
\begin{equation} \label{EQ65}
\begin{array}{l}
\displaystyle{%
  \varepsilon_k=\frac{\hbar^2k^2}{2m} - \mu + W_0 + \frac{4\pi}{k}\int_0^{\!\infty}\!\rd r\,rW_\textrm{C}(r)\sin(kr). %
}%
\end{array}
\end{equation}
Considering that both pair and three-body interactions give
contribution into the self-consistent potential
$W_0=W_0^{(2)}+W_0^{(3)}$,
$W_\textrm{C}\!(r)=W_\textrm{C}^{(2)}\!(r)+W_C^{(3)}\!(r)$, where
\begin{align} \label{EQ66}
& W_0^{(2)}=nU_{20},&  &W_\textrm{C}^{(2)}\!(r)=-U_2(r)\rho(r),
\nonumber \\
& W_0^{(3)}=2b_1-b_2, &  &W_\textrm{C}^{(3)}\!(r)=a_1\!(r)-2\rho(r)a_2\!(r).
\end{align}
Here, $U_2(r)$ is the potential of the pair interaction, %
$U_{20}=\int\! U_2(r)\rd{\bf r}$, $n=2\rho(0)$ is the particle
number density, and quantities $a_1(r)$, $a_2(r)$, $b_1$, $b_2$ for
the potential (\ref{EQ50}) are defined by the formulae (\ref{EQ61}). %
The one-particle density matrix has the form
\begin{equation} \label{EQ67}
\begin{array}{l}
\displaystyle{%
  \rho(r)=\frac{1}{2\pi^2r}\int_0^{\!\infty}\!\!\rd k\,k\sin(kr) f(\varepsilon_k), %
}%
\end{array}
\end{equation}
where $f(\varepsilon_k)=\big[\exp(\beta\varepsilon_k) +1\big]^{-1}$. %

At low temperatures in a degenerate system, its properties are
determined by quasiparticles located near the Fermi surface. %
In this case, the notion of the effective mass of a quasiparticle
$m_*$ can be introduced. We define the Fermi wave number at finite
temperatures by the relation
\begin{equation} \label{EQ69}
  \frac{\hbar^2k_{\rm F}^2}{2m} - \mu + W_0 + \frac{4\pi}{k_{\rm F}}\int_0^{\!\infty}\!\rd r\,rW_\textrm{C}(r)\sin(k_{\rm F}r) =0.
\end{equation}
Then, near the Fermi surface $k=k_{\rm F}+\Delta k$, so that the
dispersion law of a quasiparticle can be represented in the form
$\varepsilon_k=(\hbar^2k_{\rm F}/m_*)\Delta k$, where the effective mass %
is defined by the formula
\begin{equation} \label{EQ70}
  \frac{1}{m_*}=\frac{1}{m} - \frac{4\pi}{k_{\rm F}\hbar^2}\int_0^{\!\infty}\!\rd r\,r^3W_\textrm{C}(r)j_1(k_{\rm F}r), %
\end{equation}
and $j_1(x)=(\sin x-x\cos x)/x^2$  is the first order spherical
Bessel function. It is seen that the effective mass is determined by
the exchange part of the self-consistent potential. %
It should be stressed that within the self-consistent field model,
the formula (\ref{EQ70}) is exact, and it is fair for any densities
at which the use of the self-consistent field approximation is still
permissible. Within the scope of the model itself, the constraint on
density cannot be derived. For estimation of the limiting value of
density at which the self-consistent field model remains true, one
should calculate a correction due to the correlation Hamiltonian
(\ref{EQ10}) by the perturbation theory and requires it to be small
relative to the main approximation \cite{P4}. %
The contribution of three-body interactions into the effective mass
of a quasiparticle with different choices of the model potential
will be studied in detail in a separate paper. In the approximation
of the effective mass, the distribution function acquires the form
\begin{equation} \label{EQ71}
  f_k=\frac{1}{\exp\!\left[\beta\displaystyle{\frac{\hbar^2k_{\rm F}}{m_*}}(k-k_{\rm F})\right]+1 } \approx %
      \frac{1}{\exp\!\left[\beta\displaystyle{\frac{\hbar^2}{2m_*}}\big(k^2-k_{\rm F}^2\big)\right]+1 }\,. %
\end{equation}
For degenerate systems, both representations of the distribution
function (\ref{EQ71}) are equivalent. The only difference between
the functions (\ref{EQ71}) and the distribution function of an ideal
gas is the dependence of the effective mass entering (\ref{EQ71}) on
temperature and density.

The effective mass can be introduced for the non-degenerate system
as well, when the main contribution into the thermodynamics is given
by particles with small momenta. Having used the expansion
$\sin\!kr\approx kr-(kr)^3/3!$, from (\ref{EQ65}), we find the
dispersion law of quasiparticles in the non-degenerate case:
$\varepsilon_k=\hbar^2k^2/2m_{*\textrm{c}}-\mu_*$, %
where the effective mass $m_{*\textrm{c}}$ and the chemical potential $\mu_*$
are now determined by the formulae:
\begin{eqnarray} \label{EQ72}
\frac{1}{m_{*\textrm{c}}}&=&\frac{1}{m} - \frac{4\pi}{3\hbar^2}\int_0^{\!\infty}\!\rd r\,r^4W_\textrm{C}(r), \nonumber \\
\mu_*&=&\mu-W_0-4\pi\int_0^{\!\infty}\!\rd r\,r^2W_\textrm{C}(r).
\end{eqnarray}
Note that the formula (\ref{EQ70}) for the effective mass turns into
the formula (\ref{EQ72}), if $j_1(k_{\rm F}r)\approx k_{\rm F}r/3$
is put in (\ref{EQ70}).

The thermodynamic potential per unit volume, with account of
three-body interactions depending on pairs of distances between
particles and in the approximation of the effective mass, according
to (\ref{EQ38}), (\ref{EQ39}) has the form
\begin{eqnarray} \label{EQ73}
\frac{\Omega}{V}&\!\!\!=\!\!\!& -\frac{2T}{\Lambda^3}\Phi_{5/2}\!\left(\beta\frac{\hbar^2k_{\rm F}^2}{2m_*}\right) + 4\pi\!\left[\int_0^{\!\infty}\!\! U_2(r)\rho^2(r)r^2\rd r -2\rho^2(0)\!\int_0^{\!\infty}\!\! U_2(r)r^2\rd r\right] \nonumber \\
&& +\,16\pi^2 \left[ \frac{8}{3}\rho(0)\int_0^{\!\infty}\!\! r^2\rd r\!\int_0^{\!\infty}\!\! U_3(r,r')\rho^2(r')r'^2\rd r' + \frac{4}{3}\rho(0)\int_0^{\!\infty}\!\! \rho^2(r)r^2\rd r\!\int_0^{\!\infty}\!\! U_{30}(r,r';r')r'^2\rd r' \right.
\nonumber \\
&& \left.-\,\frac{8}{3}\rho^3(0)\int_0^{\!\infty}\!\! r^2\rd r\! \int_0^{\!\infty}\!\! U_3(r,r')r'^2\rd r' - \frac{4}{3}\int_0^{\!\infty}\!\! \rho(r)r^2\rd r\! \int_0^{\!\infty}\!\! U_3(r,r')\rho(r')\rho_0(r,r')r'^2\rd r' \right]\!.
\end{eqnarray}
Here, $\Lambda\!\equiv\!\big(2\pi\hbar^2/m_*T\big)^{1/2}$ is the
thermal wavelength. The first term contains one of the integrals of
the Fermi–Dirac function tabulated in the McDougall and Stoner
paper~\cite{DS}
\begin{equation} \label{EQ74}
\begin{array}{l}
\displaystyle{%
  \Phi_s(t)=\frac{1}{\Gamma(s)}\int_0^{\!\infty} \frac{z^{s-1}\,\rd z}{e^{z-t}+1} %
}%
\end{array}
\end{equation}
[$\Gamma(s)$ is the gamma function], by which all the
thermodynamical quantities of an ideal Fermi gas can be expressed.
The first term in (\ref{EQ73}) gives the contribution of a gas of
non-interacting quasiparticles. Its difference from the
thermodynamical potential of an ideal Fermi gas consists only in the
replacement of mass by the effective mass. The rest terms in
(\ref{EQ73}) are specific for the self-consistent field model and
are determined by pair and three-body interactions. Since the pressure
is associated with the thermodynamical potential by the known
relation $p=-\Omega\big/V$, then the formula (\ref{EQ73})
determines, except for sign, also the pressure. The formula for the
particle number density
\begin{equation} \label{EQ75}
\begin{array}{l}
\displaystyle{%
  n=\frac{2}{\Lambda^3}\,\Phi_{3/2}\!\left(\beta\frac{\hbar^2k_{\rm F}^2}{2m_*}\right), %
}%
\end{array}
\end{equation}
together with (\ref{EQ73}), determines the pressure as a function of
density, that is the equation of state.

\section{Fermi system at zero temperature with three-body interactions in the
model of ``semi-transparent sphere''} %

The system of Fermi particles at zero temperature can be
investigated in detail if the three-body interaction is chosen in
the form (\ref{EQ53}) which leads to somewhat more simple formulae
than for the potential (\ref{EQ50}). For the potential (\ref{EQ53}),
the parameters determining the self-consistent potential
(\ref{EQ47}) are given by the formulae
\begin{align} \label{EQ76}%
& a_1\!(r)=4\pi\sum_{l=0}^\infty\frac{1}{2l+1}  \int_0^{\!\infty}\!\! \underbar{\it{U}}_{\,3l}(r,r')\rho_l(r,r')\rho(r')r'^2\rd r', &
& a_2\!(r)=4\pi\rho(0)\!\int_0^{\!\infty}\!\! \underbar{\it{U}}_{\,30}(r,r')r'^2\rd r',  \nonumber \\
& b_1=16\pi^2\rho^2(0) \int_0^{\!\infty}\!\! r^2\rd r \!\int_0^{\!\infty}\!\! \underbar{\it{U}}_{\,30}(r,r')r'^2\rd r', &
& b_2=16\pi^2 \int_0^{\!\infty}\!\! r^2\rd r \!\int_0^{\!\infty}\!\! \underbar{\it{U}}_{\,30}(r,r')\rho^2(r')r'^2\rd r'.
\end{align}
Here,
\begin{eqnarray} \label{EQ77}
&&U_3(r,r',|{\bf r}-{\bf r}'|)=\sum_{l=0}^\infty \underbar{\it{U}}_{\,3l}(r,r')P_l(\cos\theta), %
\nonumber \\
&&\underbar{\it{U}}_{\,3l}(r,r')=\frac{2l+1}{2}\int_{-1}^{1}
U_3\!\big(r,r',\!\sqrt{r^2+r'^2-2rr'x}\,\big)\!P_l(x)\,\rd x.
\end{eqnarray}
In particular for the interaction potential of ``semi-transparent
sphere'' type (\ref{EQ57}), we have:
\begin{equation} \label{EQ78}
 \underbar{\it{U}}_{\,3l}(r,r')=U_{3\rm m}\,\theta(r_3-r)\theta(r_3-r')(2l+1)
 \bigg[\theta(r_3-r-r')\delta_{l0}+\theta(r+r'-r_3)\frac{1}{2}\int_{x_0}^{1}\!P_l(x)\rd x \bigg],%
\end{equation}
where $x_0=(r^2+r'^2-r_3^2)\big/2rr'$. At zero temperature
$f_k=\theta(k_{\rm F}-k)$ and
\begin{equation} \label{EQ79}
\begin{array}{l}
\displaystyle{%
  \rho(r)=\frac{k_{\rm F}^2}{2\pi^2r}j_1(k_{\rm F}r), \qquad
  \rho(0)=\frac{n}{2}=\frac{k_{\rm F}^3}{6\pi^2}, \qquad
  \rho_l(r,r')=\frac{2l+1}{2\pi^2}\int_0^{k_{\rm F}}\!\!\rd k\,k^2 j_l(kr)j_l(kr'). %
}%
\end{array}
\end{equation}
Taking into account the latter relations, we find
\begin{eqnarray} \label{EQ80}
a_1\!(r)&\!\!\!=\!\!\!&U_{3\rm m}\,\theta(r_3-r)\frac{k_{\rm F}^3}{2\pi^3}\,\frac{1}{k_{\rm F}r}\!\left[ j_0^2(k_{\rm F}r_3)- j_0(k_{\rm F}r_3)j_0[k_{\rm F}(r_3-r)] + \int_{\!-k_{\rm F}(r_3-r)}^{k_{\rm F}r_3} \rd y\,j_0(k_{\rm F}r-y)j_1(y)  \right],
\nonumber \\
a_2\!(r)&\!\!\!=\!\!\!&\frac{U_{3\rm m}}{72\pi}(k_{\rm F}r_3)^3\,\theta(r_3-r) \!\left[\left(\frac{r}{r_3}\right)^{\!3}-12\left(\frac{r}{r_3}\right) + 16 \right],
\nonumber \\
b_1&\!\!\!=\!\!\!&\frac{5\pi^2}{6}\rho^2(0)U_{3\rm m}r_3^6 = \frac{5}{216\pi^2}U_{3\rm m}(k_{\rm F}r_3)^6=0.0023\,U_{3\rm m}(k_{\rm F}r_3)^6,
\nonumber \\
b_2&\!\!\!=\!\!\!&\frac{U_{3\rm m}}{12\pi^2}\Big[B_3(k_{\rm F}r_3)-12(k_{\rm F}r_3)^2B_1(k_{\rm F}r_3)+16(k_{\rm F}r_3)^3B_0(k_{\rm F}r_3) \Big],
\end{eqnarray}
where $B_n(z)\equiv\int_0^{z}y^nj_1^2(y)\rd y$. When calculating
$a_1\!(r)$, the summation formula is used (see \cite{VM}, p.~133) %
\begin{equation} \label{EQ81}
  \sum_{l=0}^\infty (2l+1)j_l(u)j_l(v)P_l(x) =
  j_0\!\left(\!\sqrt{u^2+v^2-2uvx}\,\right).
\end{equation}
In the limit $k_{\rm F}r_3\ll 1$, we have more simple formulae %
\begin{equation} \label{EQ82}
  b_2\approx\frac{5}{216\pi^2}U_{3\rm m}(k_{\rm F}r_3)^6, \qquad %
  a_1\!(r)\approx  \frac{U_{3\rm m}}{432\pi^3r_3^3}(k_{\rm F}r_3)^6\,\theta(r_3-r) %
                   \!\left[\left(\frac{r}{r_3}\right)^{\!3}-12\left(\frac{r}{r_3}\right) + 16 \right].
\end{equation}
Note that even at $k_{\rm F}r_3=1$, the formulae (\ref{EQ82}) give a
mistake of the order and less than 10\%. For the following, it is
convenient to define a characteristic value of density through the
radius of the pair interaction:
\begin{equation} \label{EQ83}
\begin{array}{ll}
\displaystyle{\hspace{0mm}%
  \frac{1}{n_*}\equiv \frac{4\pi}{3}r_2^3.
}%
\end{array}
\end{equation}
If, for example, $r_2=3\!\times\!10^{-8}$~{cm} is taken,
then $n_*=0.88\!\times\!10^{22}$~{cm}$^{-3}$.
This density is close to the particle number density in liquid.
Note that, for example, the particle number density in liquid
helium-3 at near-zero temperature and vapor pressure:
$n_{\rm{^3He}}\approx 1.6\!\times\!10^{22}$~{cm}$^{-3}$.
Since $k_{\rm F}r_3=(9\pi/4)^{1/3}(r_3/r_2)(n/n_*)^{1/3} $,
then the condition $k_{\rm F}r_3\ll 1$ can be written in the form
\begin{equation} \label{EQ84}
\begin{array}{ll}
\displaystyle{\hspace{0mm}%
  (r_3/r_2)(n/n_*)^{1/3} \ll 1.
}%
\end{array}
\end{equation}
Since it should be considered likely that $r_3 \sim r_2$, %
the noted condition is satisfied in the limit of low densities. %
However, as was noted, even at $n \sim n_*$ the formulae
(\ref{EQ82}) are valid with good accuracy. Let us introduce the
designation that will be used below for the dimensionless density
\begin{equation} \label{EQ85}
\begin{array}{ll}
\displaystyle{\hspace{0mm}%
  \tilde{n}\equiv\frac{n}{n_*}=\left(\frac{r_2}{r_S}\right)^{\!3}, %
}%
\end{array}
\end{equation}
where ${r_\textrm{S}=\left[{3}/({4\pi n})\right]^{1/3}}$ %
is the radius of a sphere whose volume equals the volume per particle. %
It is important to note that, even in the case when the
dimensionless density (\ref{EQ85}) is considerably less than unity,
the value of the dimensional density proves to be much greater than
that of low-density gases. In the following formulae we will
also use the next designations:
\begin{equation} \label{EQ86}
  \lambda\equiv\frac{r_3}{r_2},\qquad %
  \alpha\equiv\frac{a_0}{r_2},\qquad %
  u\equiv\frac{U_{3\rm m}}{U_{2\rm m}}, %
\end{equation}
$a_0= mU_{2m}r_{2}^3\big/3\hbar^2$. With account of (\ref{EQ66}),
the relations (\ref{EQ79}), (\ref{EQ80}) determine the contribution
of three-body interactions into the direct and exchange parts of the
self-consistent potential
\begin{equation} \label{EQ87}
   W_0^{(3)}=2b_1-b_2,\qquad W_\textrm{C}^{(3)}\!(r)=a_1\!(r)-3n\frac{j_1(k_{\rm F}r)}{k_{\rm F}r}a_2\!(r).
\end{equation}
The contribution of pair forces into these quantities for the
potential of ``semi-transparent sphere'' type (\ref{EQ55}) is given
by the formulas
\begin{equation} \label{EQ88}
 W_0^{(2)}=U_{2\rm m}\tilde{n},\qquad
 W_\textrm{C}^{(2)}\!(r)=\left\{
               \begin{array}{ll}
                 -U_{2\rm m}\displaystyle{\frac{3}{2}\,n\,\frac{j_1(k_{\rm F}r)}{k_{\rm F}r}}, & \quad r<r_2, \\
                 \quad 0, & \quad r>r_2.
               \end{array} \right.
\end{equation}

The obtained relations (\ref{EQ87}), (\ref{EQ88}) for the
self-consistent potential enable us to determine by the formula
(\ref{EQ70}) the contribution of pair and three-body interactions
into the effective mass. With account of only pair forces, the
effective mass is determined by the expression
\begin{equation} \label{EQ89}
  \frac{m}{m_{*2}}=1+\frac{6}{\pi}\frac{\alpha}{(k_{\rm F}r_2)^2}\,B_2(k_{\rm F}r_2). %
\end{equation}
Hence, it follows that for the repulsive pair interaction $U_{2\rm m}>0$, %
the effective mass of a quasiparticle proves to be less than the
mass of a free particle, and for the attraction $U_{2\rm m}<0$, it is
greater. The effective mass with account of three-body interactions
in this case has the form
\begin{equation} \label{EQ90}
\begin{array}{l}
\displaystyle{%
  \frac{m}{m_{*}}=\frac{m}{m_{*2}}-\frac{4\pi m}{k_{\rm F}\hbar^2}\int_0^{r_3}\!\!a_1\!(r)j_1(k_{\rm F}r)r^3\rd r + %
  \frac{12\pi m n}{k_{\rm F}^2\hbar^2}\int_0^{r_3}\!\! j_1^2(k_{\rm F}r)a_2\!(r)r^2\rd r. %
}%
\end{array}
\end{equation}
It should be stressed that, for all densities at which the
self-consistent field approximation holds, the formula (\ref{EQ90})
is exact. At low relative densities, when the condition (\ref{EQ84})
is satisfied, the formula for the total effective mass (\ref{EQ90})
can be represented in the form of expansion in powers of density
\begin{eqnarray} \label{EQ91}
  \frac{m}{m_{*}}&=&1+\frac{3}{10}\,\alpha\,\tilde{n}
  - \frac{3}{70}\!\left(\frac{9\pi}{4}\right)^{\!2/3}\alpha\,\tilde{n}^{5/3}
  + \frac{159}{2560}\,\alpha\,u\,\lambda^8\,\tilde{n}^2 \nonumber\\
  &=&
  1+0.3\,\alpha\,\tilde{n} - 0.158\,\alpha\,\tilde{n}^{5/3}
  + 0.0621\,\alpha\, u\,\lambda^8\,\tilde{n}^2.
\end{eqnarray}
In (\ref{EQ89}), (\ref{EQ91}) and below we use the designations
(\ref{EQ86}). The second and the third terms in (\ref{EQ91}) are
determined by pair forces, and the fourth by three-body forces. As
it is seen, with increasing density the contribution of three-body
forces into the effective mass rises faster than that of pair
forces, and the relative role of three-body forces rises.
Furthermore, attention is drawn to a strong dependence (as $\lambda^8$) of
the contribution of three-body forces on the ratio of the radii of
three-body and pair interactions.

\begin{figure}[t!]
\centering %
\includegraphics[width = 0.5\textwidth]{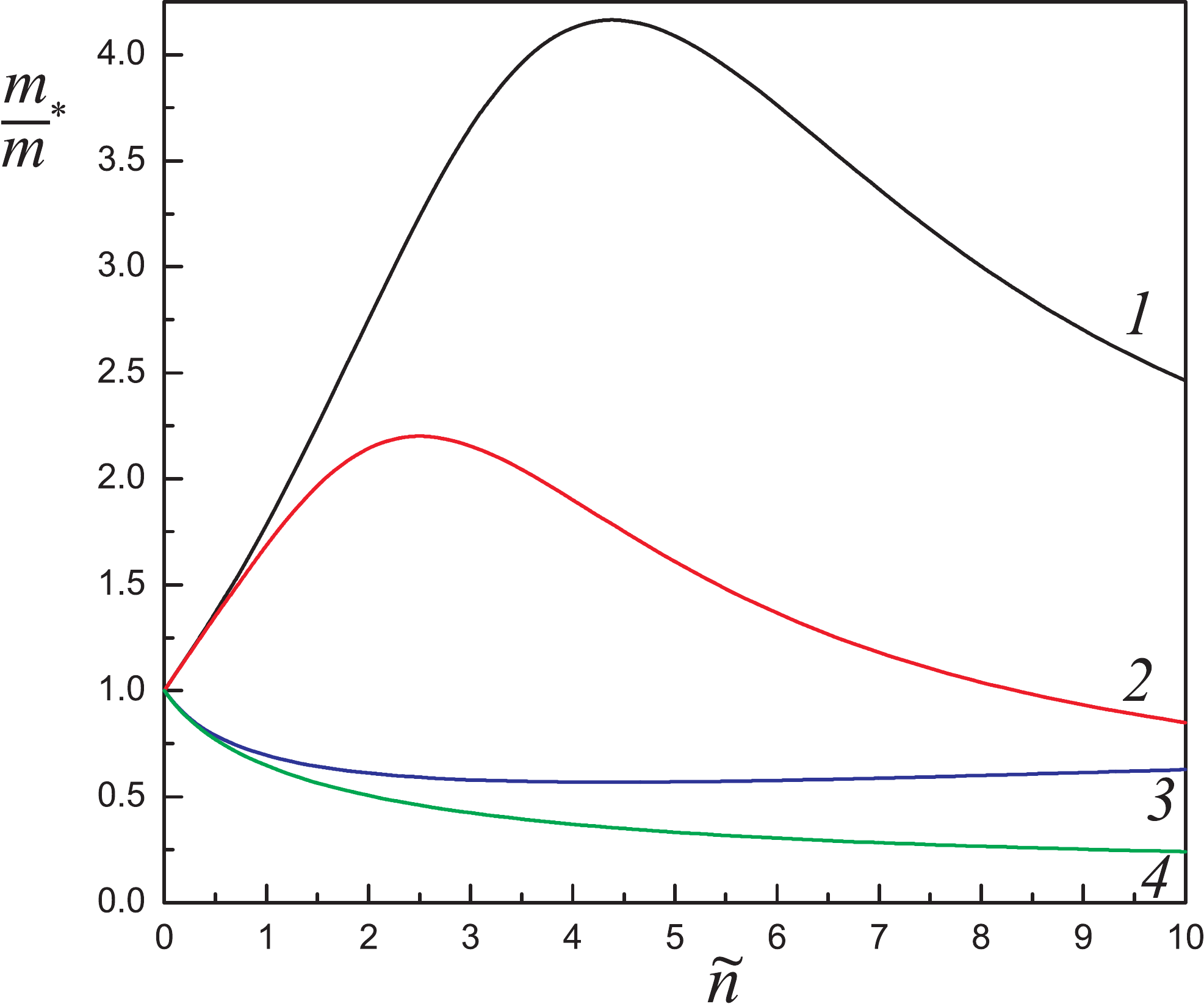} %
\caption{\label{fig01} %
(Color online) Dependencies of the effective mass on density: %
(1) $\alpha=-2.5$, $u=0$; (2) $\alpha=-2.5$, $u=-0.3$;
(3) $\alpha=2.5$, $u=0$; (4) $\alpha=2.5$, $u=1$. %
It is everywhere assumed that $\lambda=1$.
}%
\end{figure}

Dependencies of the effective mass on density, calculated by the
exact formula (\ref{EQ90}), are shown in figure~\ref{fig01}. For the
attractive interaction and with neglect of three-body forces, the
effective mass proves to be greater than the mass of a free particle and
monotonously rises with an increasing density, reaching its maximum. With
a further increase of density, the rise of the effective mass changes
into its fall, although, as before for all physically reasonable
densities, it remains greater than the mass of a free particle (curve~1
in figure~\ref{fig01}). Accounting for the three-body interaction with a
positive constant leads to a reduction of the region of the rise of the
effective mass and to its faster  decrease at high densities
(curve~2 in figure~\ref{fig01}). For the repulsive interaction and in the
absence of three-body forces, the effective mass for all reasonable
densities is less than the mass of a free particle (curve~3 in figure~\ref{fig01}). Accounting for the three-body interaction with a positive
constant leads to a faster monotonous decrease of the effective
mass (curve~4 in figure~\ref{fig01}).

The total pressure $p=-\Omega\big/V$ of the Fermi system with the
three-body interaction of the form (\ref{EQ53}) at finite
temperatures, according to the general formulae
(\ref{EQ38}), (\ref{EQ39}), is as follows
\begin{eqnarray}
\label{EQi01}
p &=& p_0 + p_2 + p_3, \nonumber \\
p_0&=&\frac{2T}{\Lambda^3}\Phi_{5/2}\!\left(\beta\frac{\hbar^2k_{\rm F}^2}{2m_*}\right), \qquad  p_2= 4\pi\!\left[-\!\int_0^{\!\infty}\!\! U_2(r)\rho^2(r)r^2\rd r +2\rho^2(0)\!\int_0^{\!\infty}\!\! U_2(r)r^2\rd r\right],
\nonumber \\
p_3 &=& \!\int\!\! \rd {\bf r}\rd {\bf r}'\, U_3(r,r',|{\bf r}-{\bf r}'|) %
  \!\bigg[\! -4\rho(0)\rho^2(r')+\frac{8}{3}\rho^3(0)+\frac{4}{3}\rho(r)\rho(r')\rho(|{\bf r}-{\bf r}'|)\bigg], %
\end{eqnarray}
where the term $p_0$ is a contribution of a gas of fermions with the
effective mass $m_*$, and the terms $p_2$ and $p_3$ give,
respectively, contributions of the pair and three-body interactions. %
Together with the formula (\ref{EQ75}) for the particle number
density, formulae (\ref{EQi01}) define in a general form the system's
equation of state for the considered form (\ref{EQ53}) of three-body
interactions.

At zero temperature, the pressure of the fermion gas
\begin{equation} \label{EQ92}
\begin{array}{ll}
\displaystyle{\hspace{0mm}%
  p_0= \frac{(3\pi^2)^{2/3}}{5}\frac{\hbar^2}{m_*}n^{5/3}, %
}%
\end{array}
\end{equation}
and contributions of the pair and three-body forces into the total
pressure (\ref{EQi01}) in the model of ``semi-transparent sphere''
potentials for them [formulae (\ref{EQ55}), (\ref{EQ57})], %
with the expansion formulae (\ref{EQ76})--(\ref{EQ79}) taken
into account, acquire the form
\begin{eqnarray} \label{EQ93}
  p_2&=& -\frac{U_{2\rm m}n_*}{2}\bigg[ \frac{6}{\pi}\,\tilde{n}\,B_0(k_{\rm F}r_2) - \tilde{n}^2\bigg], \\
\label{EQ95}
  p_3&=& 4\rho(0)\bigg(\frac{2}{3}b_1-b_2\bigg) +
       \frac{16\pi}{3} \int_0^{\!\infty}\!\! \rho(r)a_1\!(r)r^2\rd r,
\end{eqnarray}
where $b_1$, $b_2$, $a_1(r)$ are defined by the formulae (\ref{EQ80}).
In the case of densities $n\leqslant n_*$, using the expansion
\[
{B_n(z)\approx \frac{z^{n+3}}{9(n+3)}-\frac{z^{n+5}}{45(n+5)}+\frac{z^{n+7}}{525(n+7)}}
\]
we find the dependence of the pressure due to the pair interaction
on density
\begin{eqnarray} \label{EQ96}
\frac{p_2}{p_{0*}}&= &\frac{5\,\alpha}{(12\pi^2)^{1/3}}
\bigg[ \tilde{n}^2 +\frac{3}{25}\!\left(\frac{9\pi}{4}\right)^{\!2/3}\tilde{n}^{8/3}
-\frac{81\pi}{4900}\!\left(\frac{9\pi}{4}\right)^{\!1/3}\tilde{n}^{10/3} \bigg]\nonumber\\
&=& \frac{5\,\alpha}{(12\pi^2)^{1/3}} \bigg[ \tilde{n}^2 +0.442\,\tilde{n}^{8/3}
-0.0997\,\tilde{n}^{10/3} \bigg].
\end{eqnarray}
Here the pressure is related to that of a gas of particles at the
characteristic density (\ref{EQ83}): $p_{0*}=$ \linebreak  $\frac{1}{5}{(3\pi^2)^{2/3}}({\hbar^2}/{m})n_*^{5/3}$. %
The sign of pressure (\ref{EQ96}) is determined by the sign of the
pair interaction constant and, at negative value of this constant, the
interaction contribution into the pressure is negative. The main
term in the expansion of pressure due to the three-body interaction
(\ref{EQ95}) has the form
\begin{equation} \label{EQ97}
  \frac{p_3}{p_{0*}}= \frac{477}{2560}\,\alpha\,u\,\lambda^8\,\tilde{n}^{11/3}=  %
                      0.186\,\alpha\,u\,\lambda^8\,\tilde{n}^{11/3}.
\end{equation}
Attention should be paid to the fact that the expansion of $p_2$ holds in even
powers of the quantity $\tilde{n}^{1/3}$ and the expansion of $p_3$
in odd powers, and the latter begins with a high power.

\begin{figure}[!t]
\centering
\includegraphics[width = 0.5\textwidth]{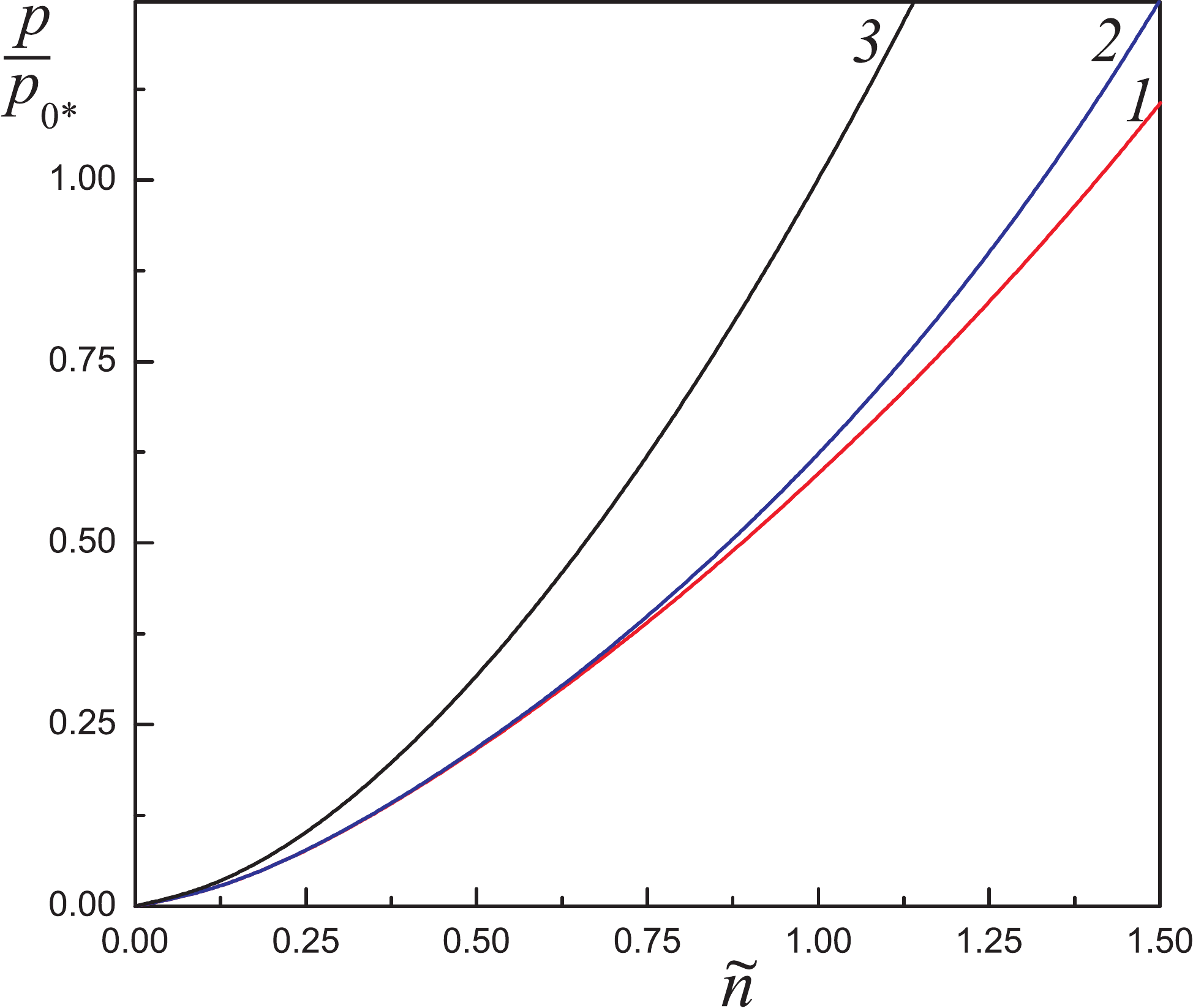} %
\caption{\label{fig02} %
(Color online) Comparison of the accurate dependence of the pressure (1),
calculated by the formulae (\ref{EQ92})--(\ref{EQ95}) %
and the approximate dependence (2), calculated by the formula (\ref{EQ99}), %
at $\alpha=-0.3$, $u=-1$, $\lambda=1$. (3) Ideal Fermi gas. %
}%
\end{figure}

Since the effective mass depends on density, then the pressure of a
gas of quasiparticles (\ref{EQ92}), with account of (\ref{EQ91}),
can also be represented in the form of expansion in powers of
density:
\begin{eqnarray} \label{EQ98}
\frac{p_0}{p_{0*}}&=& \tilde{n}^{5/3}+\frac{3}{10}\,\alpha\,\tilde{n}^{8/3}
-\frac{3}{70}\!\left(\frac{9\pi}{4}\right)^{\!2/3}\alpha\,\tilde{n}^{10/3}
+\frac{159}{2560}\,\alpha\,u\,\lambda^8\,\tilde{n}^{11/3}
\nonumber\\
&=&\tilde{n}^{5/3}+0.3\,\alpha\,\tilde{n}^{8/3}
-0.158\,\alpha\,\tilde{n}^{10/3}+0.0621\,\alpha\,u\,\lambda^8\,\tilde{n}^{11/3}.
\end{eqnarray}
Taking into account (\ref{EQ96})--(\ref{EQ98}), we find the
expansion of the total pressure in powers of density:
\begin{eqnarray} \label{EQ99}
\frac{p}{p_{0*}}&=& \tilde{n}^{5/3}+\frac{5}{(12\pi^2)^{1/3}}\,\alpha\,\tilde{n}^2
  +\frac{3}{4}\,\alpha\,\tilde{n}^{8/3}
  -\frac{207}{1960}\!\left(\frac{3\pi^2}{2}\right)^{\!1/3}\alpha\,\tilde{n}^{10/3}
  +\frac{159}{640}\,\alpha\,u\,\lambda^8\,\tilde{n}^{11/3}
\nonumber \\
&=& \tilde{n}^{5/3}+1.018\,\alpha\,\tilde{n}^2
  +0.75\,\alpha\,\tilde{n}^{8/3}
  -0.259\,\alpha\,\tilde{n}^{10/3}
  +0.248\,\alpha\,u\,\lambda^8\,\tilde{n}^{11/3}.
\end{eqnarray}
Comparison of the accurate dependence of the total pressure on
density, calculated by the formulae (\ref{EQ92})--(\ref{EQ95})
and the approximate dependence, calculated by the formula
(\ref{EQ99}), is given in figure~\ref{fig02}. It is seen that even at
$\tilde{n}\sim 1$, the pressure, calculated by the approximate
formula, differs weakly from its accurate value. 

Let us discuss the issue of the thermodynamic stability of the
Fermi system, assuming $\tilde{n}< 1$. For a system to be stable, it
is necessary that its compressibility or (which is the same) the
squared speed of sound should be positive, so that $\partial p/\partial n >0$. %
Since in this case the two first terms give the main contribution
into the total pressure (\ref{EQ99}) and qualitatively taking into account
 the new effect of three-body interactions, %
we find the condition of stability in the form
\begin{equation} \label{EQ100}
  1+\frac{9}{2}\!\left(\frac{4}{9\pi}\right)^{\!2/3}\alpha\,\tilde{n}^{1/3} %
   +\frac{1749}{3200}\,\alpha\,u\,\lambda^8\,\tilde{n}^2 > 0. %
\end{equation}
For positive constants of pair and three-body interactions, the
system is always stable. More interesting is the case when the
constant of the pair interaction is negative. Then, without account
of three-body interactions, the condition of stability should be
satisfied which can be represented in equivalent forms
\begin{equation} \label{EQ101}
  \tilde{n}^{1/3} < \left(\frac{\pi^2}{18}\right)^{\!1/3}|\alpha|^{-1}\approx 0.82\,|\alpha|^{-1},\qquad %
  r_\textrm{S} > \left(\frac{18}{\pi^2}\right)^{\!1/3}|a_0|\approx 1.22\,|a_0|. %
\end{equation}
Accounting for the three-body interaction, if it is of a repulsion
character, extends the region of stability and can lead to
stabilization of the system with the negative pair potential %
at arbitrary densities in case of fulfilment of the following (\ref{EQ100}) condition
\begin{equation} \label{EQ102}
\begin{array}{ll}
\displaystyle{\hspace{0mm}%
  |u|> \frac{200}{583\,\pi^4}\frac{5^5}{3^3}\frac{|\alpha|^5}{\lambda^8} \approx %
       0.408 \frac{|\alpha|^5}{\lambda^8}.
}%
\end{array}
\end{equation}

\begin{figure}[b!]
\centering
\includegraphics[width = 0.5\textwidth]{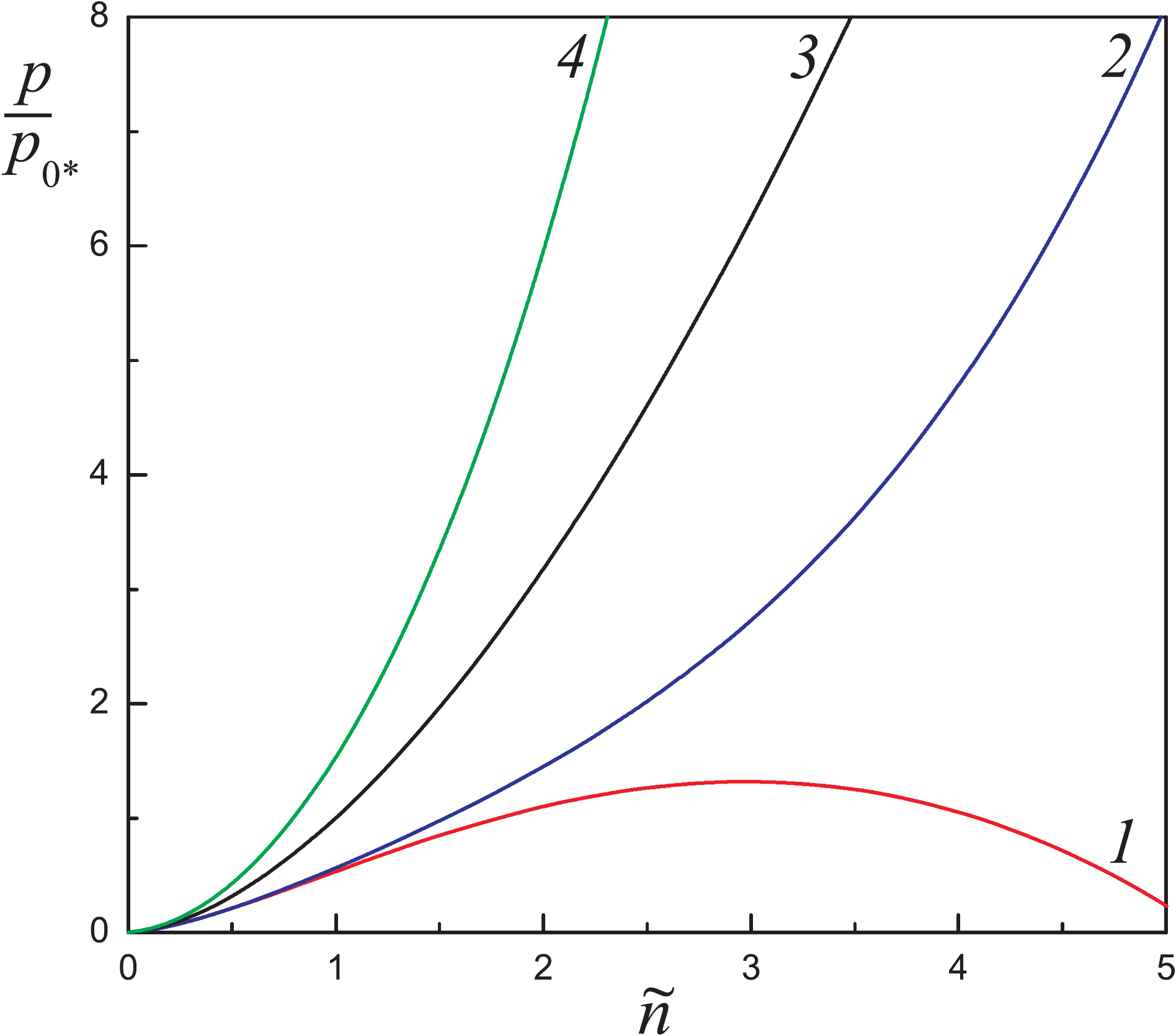} %
\caption{\label{fig03} %
(Color online) Dependencies of the pressure on density: %
(1) $\alpha=-0.3$, $u=0$; (2) $\alpha=-0.3$, $u=-0.5$; %
(3) an ideal Fermi gas; (4) $\alpha=0.3$, $u=1$. %
It is everywhere assumed that $\lambda=1$.
}%
\end{figure}

Some dependencies of the pressure on density for different signs of
the interparticle interaction are shown in figure~\ref{fig03}. In the case of
attraction and with neglect of three-body interactions, the
spatially uniform state is stable only at low densities for which
the condition (\ref{EQ101}) is satisfied. At high densities, the
pressure decreases with an increase of density (curve~1 in figure~\ref{fig03}) and
the spatially uniform state ceases to be stable. Sufficiently
intensive three-body repulsive forces lead to stabilization of the
system with pair attractive forces (curve~2 in figure~\ref{fig03}).

\begin{figure}[!t]
\centering
\includegraphics[width = 0.5\textwidth]{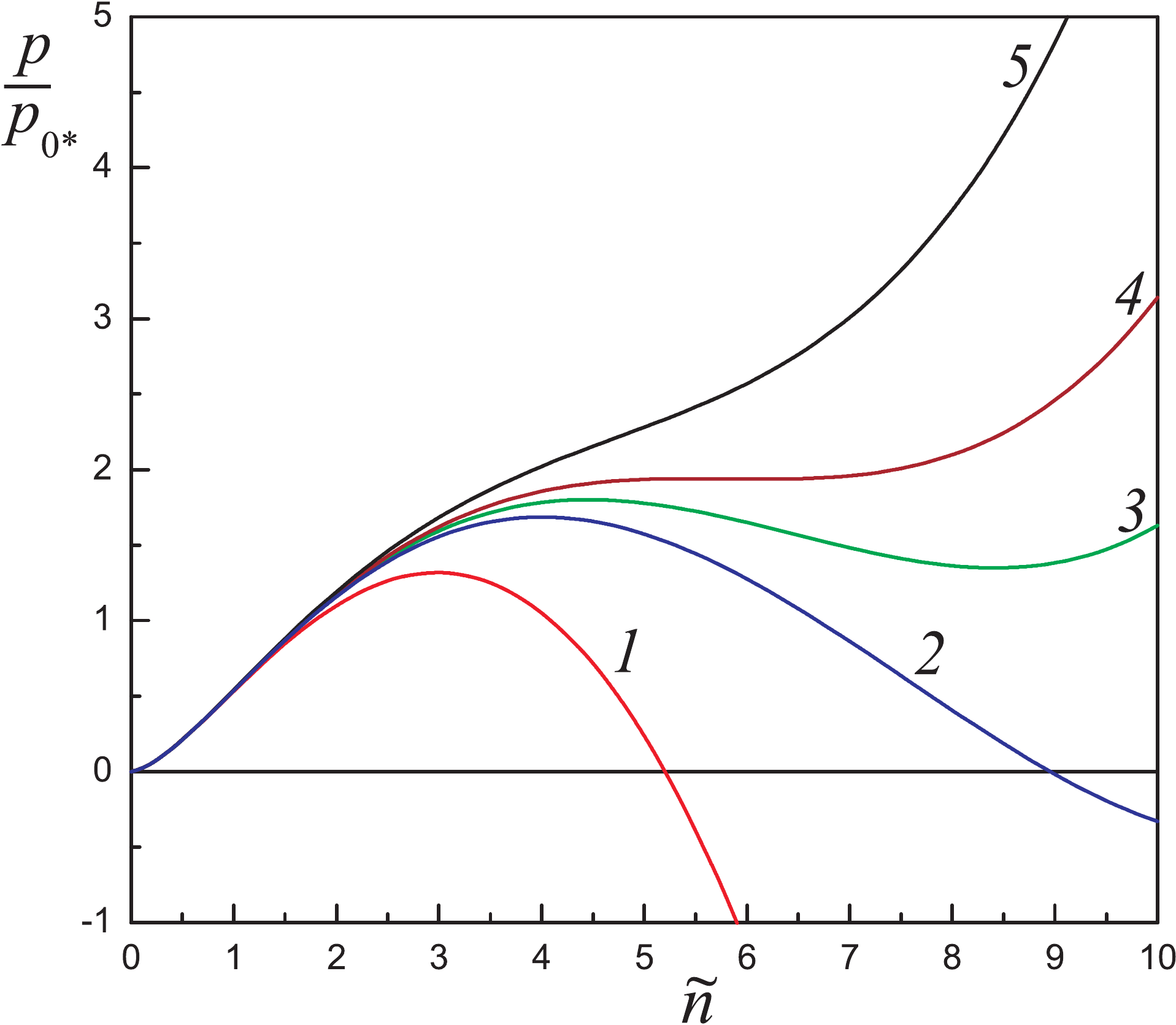} %
\caption{\label{fig04} %
(Color online) The effect of three-body repulsive forces ($U_{3\rm m}>0$) %
on stability of the Fermi system with pair attractive forces: %
(1) $u=0$; (2) $u=-0.085$; (3) $u=-0.098$; (4) $u=-0.11$; (5) $u=-0.13$. %
It is everywhere assumed that $\alpha=-0.3$ and $\lambda=1$.
}%
\end{figure}

The effect of three-body repulsive forces on the stability of the
Fermi system is illustrated more in detail in figure~\ref{fig04}. With
an increasing strength of the three-body repulsive interaction, the
regions of stability of such a system extend (curve~2, 3 in figure~\ref{fig04}), and a further growth of the intensity of the three-body
interaction leads to stabilization of the system for all physically
reasonable values of density.

\begin{figure}[!b]
\centering
\includegraphics[width = 0.5\textwidth]{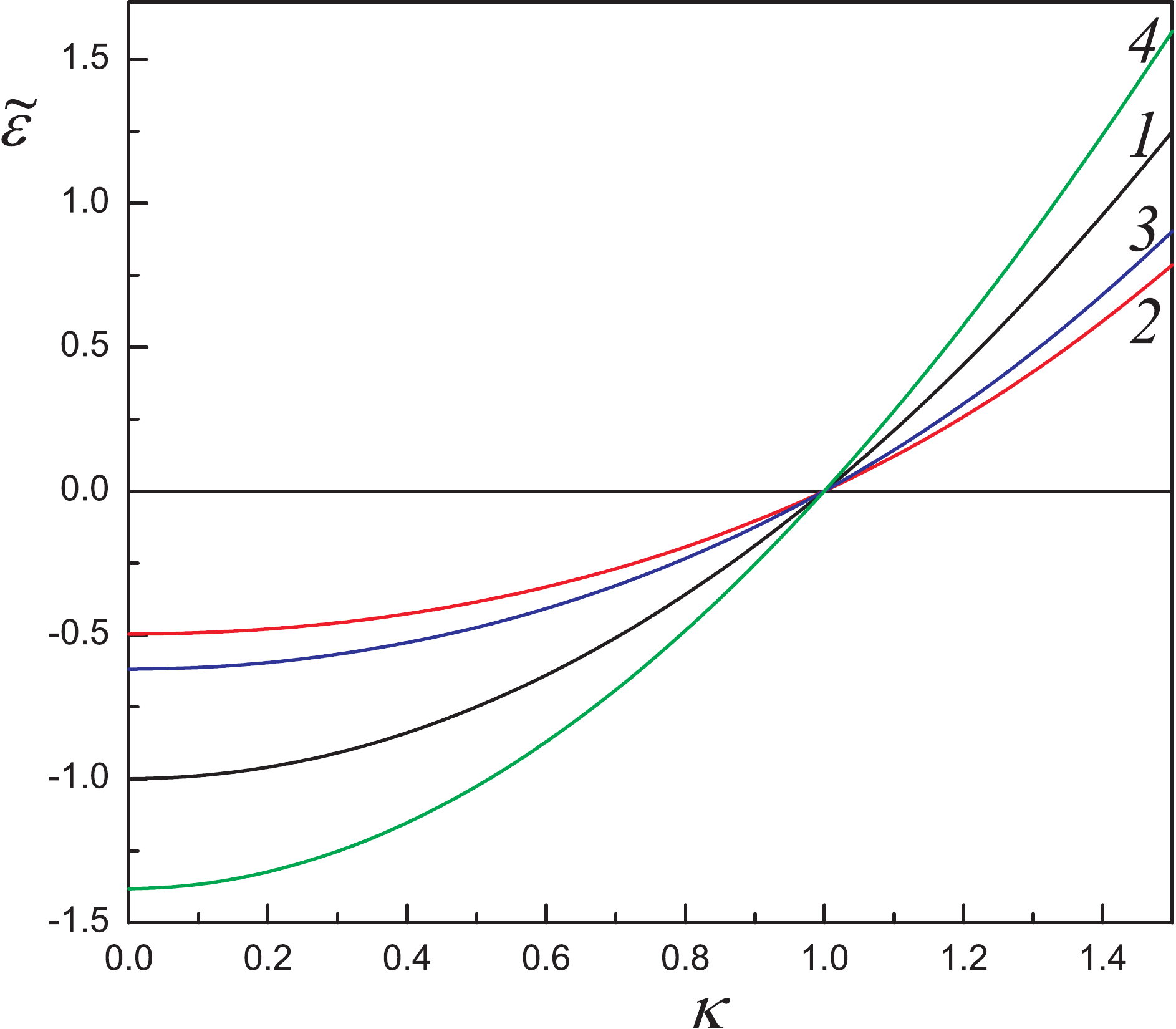}
\caption{\label{fig05}
(Color online) The quasiparticle energy spectrum $\tilde{\varepsilon}(\kappa)$
(in the variables
$\tilde{\varepsilon}\equiv\varepsilon_k/\varepsilon_{\rm F}$, $\kappa\equiv k/k_{\rm F}$):
(1) Ideal Fermi gas [$\tilde{\varepsilon}(\kappa)=\kappa^2-1$, $\tilde{m}_*\equiv m_*/m =1$];\, (2) $\alpha=-2.5$, $u=0$, $\tilde{n}=1$ ($\tilde{m}_*=1.78$);
(3) $\alpha=-2.5$, $u=-1$, $\lambda=1$, $\tilde{n}=1$ ($\tilde{m}_*=1.50$);\,
(4) $\alpha=2.5$, $u=-1$, $\lambda=1$, $\tilde{n}=1$ ($\tilde{m}_*=0.75$).
}
\end{figure}

The form of the quasiparticle energy spectrum with account of the
interparticle interactions is shown in figure~\ref{fig05}. Accounting for only
the pair attraction leads to a slowing down of the energy increase
with increasing momentum relative to an ideal Fermi gas (curve~2 in
figure~\ref{fig05}). Accounting for the repulsive three-body forces, against
the background of the attractive pair forces, additionally yields a
small increase of the quasiparticle energy (curve~3 in figure~\ref{fig05}). %
The effect of three-body forces on the spectrum is appreciably
weaker than the role of pair forces, so that accounting for the
attractive three-body forces in the presence of the pair repulsion
weakly prevents the increase of the quasiparticle energy as momentum
increases (curve~4 in figure~\ref{fig05}).

\section{Conclusion}

The self-consistent field equations are obtained, and, within this
model, thermodynamic relations are derived for a normal system
of Fermi particles with account of both pair and three-body forces.
The satisfaction of the essential requirement of fulfilment of all
thermodynamic relations already in the self-consistent approximation
leads to a unique formulation of the self-consistent field model.
Emphasized are the novelty and universality of the proposed approach
to the formulation of this model and the usefulness of the method
for describing not only Fermi, but also Bose systems, in particular
phonons, and relativistic quantized fields. The approach being
developed allows us also, as shown in the paper, to account for
many-body interactions naturally.

It is shown that three-body interactions of zero radius give no
contribution into the self-consistent field and, in order to account
for the effects due to such interactions, it is necessary to account
for their nonlocality. The case of the spatially uniform system is
considered in detail. General formulae are derived for the system's
equation of state and the effective mass of quasiparticles with
account of three-body forces. Dependencies of the quasiparticle
effective mass and the system's pressure on density at zero
temperature are obtained, with pair and three-body forces accounted
for in the model of interaction potentials of ``semi-transparent
sphere'' type.

It is shown that pair interactions of repulsive character reduce the
quasiparticle effective mass relative to the mass of a free particle
while attractive pair interactions, on the contrary, raise it. %
The effective mass and pressure are numerically calculated at zero
temperature and expansions of these quantities are derived in powers
of the relative density with account of three-body forces. %
It is shown that the relative contribution of three-body
interactions into thermodynamic quantities rises with an increasing
density. The effect of three-body forces on the stability of the
Fermi system is considered, and it is shown that in the case of
repulsion, their being taken into account extends the region of stability
and can lead to stabilization of the system with pair attraction.
The quasiparticle energy spectrum is calculated with account of the
interparticle interactions.

\ukrainianpart
\title{Модель самоузгодженого поля для фермі-систем \\ з врахуванням
тричастинкових взаємодій}
\author{Ю.М. Полуектов, О.О. Сорока, С.М. Шульга}
\address{Інститут теоретичної фізики ім. О.І. Ахієзера, ННЦ ХФТІ,
вул. Академічна, 1, 61108 Харків, Україна} %

\makeukrtitle

\begin{abstract}
\tolerance=3000%
На основі мікроскопічної моделі самоузгодженого поля побудовано
термодинаміку системи багатьох фермі-частинок при скінчених
температурах з врахуванням тричастинкових взаємодій і отримано
рівняння руху квазічастинок. Показано, що дельтаподібна
тричастинкова взаємодія не дає внеску в самоузгоджене поле, і для
опису тричастинкових сил слід враховувати їх нелокальність.
Детально розглянуто просторовооднорідну систему і в рамках
розвиненого мікроскопічного підходу отримано загальні формули для
ефективної маси ферміону і рівняння стану системи з врахуванням
внеску тричастинкових взаємодій. Для потенціалу типу
``напівпрозорої сфери'' при нулі температур чисельно розраховано
ефективну масу і тиск. Знайдено розвинення ефективної маси і тиску
за ступенями щільності. Показано, що при врахуванні тільки парних сил,
взаємодія, що має характер відштовхування, зменшує ефективну масу
квазічастинки порівняно з масою вільної частинки, а у разі
притягання~--- збільшує. Розглянуто питання термодинамічної стійкості
фермі-системи і показано, що тричастинкова взаємодія, яка має
характер відштовхування, розширює область стійкості системи з
міжчастинковим парним притяганням. Розраховано енергетичний спектр
квазічастинки з врахуванням тричастинкових сил.
\keywords самоузгоджене поле, тричастинкові взаємодії, ефективна
маса, ферміон, \\ рівняння стану
\end{abstract}

\end{document}